\documentclass[amsmath,amssymb,aps, prb]{revtex4-1}
\usepackage{dutchcal}
\usepackage{enumerate}
\usepackage{graphicx}% Include figure files
\usepackage{dcolumn}% Align table columns on decimal point
\usepackage{bm}% bold math
\usepackage{graphicx}% Include figure files
\usepackage{bbold} %for identity matrix
\usepackage{graphicx}% Include figure files
\usepackage{dcolumn}% Align table columns on decimal point
\usepackage[left=3cm,top=2.5cm,right=3cm,bottom=2.5cm]{geometry}
\usepackage{array}
\usepackage{epsfig}
\usepackage{wrapfig}
\usepackage{color}
\usepackage{soul}
\usepackage{braket}
\usepackage{verbatim}
\usepackage{mathtools}
\usepackage{mathrsfs}
\usepackage[english]{babel}
\usepackage{textcomp}
\usepackage{float}
\usepackage{comment}
\usepackage[mathscr]{euscript}
\excludecomment{toexclude}
\usepackage{changepage} %to move tables to the left using "adjustwidth"

\usepackage[left=3cm,top=2.5cm,right=3cm,bottom=2.5cm]{geometry}
\usepackage{epsfig}
\usepackage{color}
\usepackage{braket}
\usepackage{verbatim}
\usepackage{mathtools}
\usepackage{amsfonts}
\usepackage[english]{babel}
\usepackage{float}
\usepackage{comment}

\excludecomment{toexclude}
\newcommand{\mr}{\mathrm}
\newcommand{\lcal}{L} % {\cal L}
\newcommand{\z}{Z} % {\cal Z}
\newcommand{\s}{S}
\newcommand{\dd}{D}
\newcommand{\E}{E}   % \mathbb{E}

\begin{document}

%\preprint{APS/123-QED}

\title{User's guide to Monte Carlo methods for evaluating path integrals}


\author{Marise J. E. Westbroek}
\affiliation{Department of Earth Science and Engineering, Imperial College London, London SW7 2BP, United Kingdom}

\author{Peter R. King}
\affiliation{Department of Earth Science and Engineering, Imperial College London, London SW7 2BP, United Kingdom}

\author{Dimitri D. Vvedensky}
\affiliation{The Blackett Laboratory, Imperial College London, London SW7 2AZ, United Kingdom}

\author{Stephan D\"{u}rr}
\affiliation{University of Wuppertal, Gau\ss stra\ss e 20, D-42119 Wuppertal, Germany}
\affiliation{J\"ulich Supercomputing Centre, Forschungszentrum J\"ulich, D-52425 J\"{u}lich, Germany}

%\affiliation{%
% This line break forced with \textbackslash\textbackslash}%


%\affiliation{%
% This line break forced with \textbackslash\textbackslash}%


\begin{abstract}

We give an introduction to the calculation of path integrals on a lattice, with the quantum harmonic oscillator as an example. In addition to providing an explicit computational setup and corresponding pseudocode, we pay particular attention to the existence of autocorrelations and the calculation of reliable errors. The over-relaxation technique is presented as a way to counter strong autocorrelations. The simulation methods can be extended to compute observables for path integrals in other settings.
\end{abstract}

%\pacs{Valid PACS appear here}

\maketitle

\section{\label{Intro}Introduction}

Undergraduate physics students are familiar with quantum mechanics as formulated by Schr\"odinger, Heisenberg, and others in the 1920s.\cite{waerden68, brush87} Schr\"{o}dinger's equation and Heisenberg's matrix mechanics are based on Hamiltonian classical mechanics, which provides a direct connection between (classical) Poisson brackets and (quantum) commutators of observables. In 1933, Dirac \cite{dirac,dirac47} proposed an approach to quantum mechanics based on the Lagrangian, which he regarded as more fundamental than the Hamiltonian. Dirac suggested that the transition amplitude in quantum mechanics, also called the propagator, corresponds to the quantity $\exp(iS/\hbar)$, in which $S$ is the classical action evaluated along the path the particle takes.

In 1948, Feynman\cite{feynman48} extended Dirac's ideas and formulated quantum mechanics based on the sum over \emph{all} paths between fixed initial and final states. Each path contributes a pure phase $\exp(iS/\hbar)$ to the propagator, as Dirac suggested, with the amplitudes of the paths combined according to the usual quantum mechanical procedure for the superposition of amplitudes. Because the sum over paths is typically an integral over a continuum of paths, this procedure is now known as the {\it path integral method}.

Feynman derived his path integral method in a seminal paper \cite{feynman48} that laid the foundation for many formal developments and applications of path integrals in other areas of physics,\cite{brush61,hibbs} most notably, in quantum field theory,\cite{huang10} statistical mechanics,\cite{wiegel75} and stochastic dynamics.\cite{wio13} In fact, although largely unknown to the physics community at that time, the notion of the integral over paths had been introduced in the 1920s by the mathematician Norbert Wiener\cite{wiener21} for diffusion and Brownian motion. Wiener's presentation had a  similar formal structure to the Feynman path integral, though in a purely classical context.\cite{brush61,kac66}

Path integrals provide an intuitively appealing framework for interpreting many aspects of quantum mechanics. A fundamental property of path integrals is the emergence of the classical limit as $\hbar\to0$. In quantum mechanics ($\hbar\ne0$), the classical path and nearby paths contribute constructively to the path integral, and others oscillate rapidly and cancel. As $\hbar\to0$, the nearby paths oscillate rapidly and also cancel, leaving only the contribution from the classical path, which minimizes the action.\cite{landau60} Quantum mechanical paths can explore regions unavailable to the classical path, leading to phenomena such as tunneling. The double-slit experiment, which is a conceptually simple demonstration of a fundamental difference between classical and quantum physics\cite{feynman65,sawant14} is an example where the path integral provides a compelling basis for interpreting this experiment.

The path integral is an explicit expression for the probability amplitude. The actual calculation of these amplitudes depends on the problem of interest. In rare cases, such as the harmonic oscillator, the path integral can be evaluated exactly, but typically either an approximate solution is found, or a perturbative expansion is done. Mean-field theory, steepest descent, and the renormalization group are established methods for obtaining exact or approximate solutions from path integrals.\cite{amit84,ZJ}

However, there are situations when approximate solutions are ineffective. Among the best-known such example is quantum chromodynamics (QCD), the prevailing theory of hadronic matter and a component of the standard model of particle physics. In a regime where the coupling constant is small, calculations based on perturbation theory have been successful. In the strong coupling regime, however, such calculations fail, and an alternative approach is required. In this case, lattice QCD, in which the original theory is discretized on a space-time lattice, provides a framework for the non-perturbative numerical evaluation of amplitudes and matrix elements. The methodology is based on Markov chain Monte Carlo methods,\cite{Creutz, Morningstar, applications} the subject of this paper.

In the following we provide a guide to the numerical evaluation of path integrals, using the harmonic oscillator as an example. We will focus on trajectories $x(t)$ in one spatial dimension. The time $t$ is described by a lattice and takes discrete values. In addition to conceptual simplicity, this toy model has the advantage of having an exact solution, which enables the verification of the methodology. Although general descriptions of the computational procedures are available,\cite{Creutz} we provide a pedagogical description of  the implementation, methods for error analysis, and suggestions for improving the computational procedures.

The organization of our paper is as follows. The theoretical framework of our calculations is set out in Sec.~\ref{sec2}, including the derivation of the path integral and the correlation functions we will use. Our computational procedure is summarized in Sec.~\ref{sec3}, including the definition of observables, the updating algorithm, the notion of thermalization, and correlations within the sampled paths. The jackknife analysis of the variance of correlated variables is the subject of Sec.~\ref{sec4}, and the autocorrelation time of a sequence of configurations is discussed in Sec.~\ref{Autocorrelation}.  The technique of over-relaxation for reducing autocorrelation times is introduced in Sec.~\ref{Overrelaxation}. Advanced topics based on the harmonic oscillator are discussed in Sec.~\ref{sec7}.

\section{\label{sec2}Theoretical background}

The solution to the initial-value problem of the Schr\"odinger equation,
\begin{equation}
i\hbar{\partial\psi\over\partial t}=\hat{H}\psi\, ,
\label{eq2.1}
\end{equation}
can be written as
\begin{equation}
\psi(x,t)=e^{-i\hat{H}t/\hbar}\psi(x,0)\, ,
\label{eq2.2}
\end{equation}
where the exponential factor is known as the {\it evolution operator}. The exponential of an operator $\hat{O}$ is defined by the Taylor series of the exponential function:
\begin{equation}
e^{\hat{O}}=\sum_{n=0}^\infty \frac{\hat{O}^n}{n!}\, .
\end{equation}
Equation~(\ref{eq2.2}) is only a formal solution to Eq.~(\ref{eq2.1}) because obtaining an explicit solution from the evolution operator is no simpler than solving the original equation.

The connection between the evolution operator and Feynman's path integral can be made by considering the matrix elements of the evolution operator between any two initial and final position eigenstates. In Dirac's bra-ket notation\cite{SI}
\begin{equation}
\langle x_f | e^{-i\hat{H}(t_f-t_i)/\hbar} | x_i\rangle=\langle x_f,t_f| x_i,t_i\rangle\, .
\label{eq2.4}
\end{equation}
These matrix elements embody all the information about how a system with the Hamiltonian $\hat{H}$ evolves, or propagates, in time, and is known as the {\it propagator}. In particular, the evolution of the wave function is given by
\begin{align}
\psi(x_f,t_f) & =\langle x_f,t_f | \psi\rangle\\
&=\!\int\langle x_f,t_f| x_i,t_i\rangle\langle x_i,t_i|\psi\rangle\,dx_i \\
&=\!\int\langle x_f,t_f| x_i,t_i\rangle\psi(x_i,t_i)\,dx_i\, ,
\end{align}
which shows that the propagator (\ref{eq2.4}) is a type of Green function known as the fundamental solution of Eq.~(\ref{eq2.1}).

\subsection{Derivation of the path integral}

The standard derivation of the path integral from the evolution operator considers the evolution of a system over a short time $\delta t$. The method can be  demonstrated for the Hamiltonian
\begin{equation}
\hat{H}={\hat{p}^2\over2m}+V(\hat{x})\, ,
\label{eq7}
\end{equation}
of a particle of mass $m$ moving in a potential $V$, where $\hat{p}$ and $\hat{x}$ signify momentum and position operators.

The propagator to be evaluated is
\begin{align}
\langle x_f,t_i+\delta t|x_i,t_i\rangle&=\langle x_f| e^{-i\hat{H}\delta t/\hbar}|x_i\rangle\\
&=\!\int\langle x_f|p\rangle\langle p|e^{-i\hat{H}\delta t/\hbar}|x_i\rangle\,dp\, .
\label{eq8}
\end{align}
We expand the exponential to first order in $\delta t$:
\begin{equation}
\langle p|e^{-i\hat{H}\delta t/\hbar}|x_i\rangle=
\bigg\langle p\,\bigg|1-{i\hat{H}\delta t\over\hbar}+\mathcal{O}(\delta t)^2\bigg|\,x_i\bigg\rangle\, .
\label{eq9}
\end{equation}
The explicit mention of $\mathcal{O}(\delta t)^2$ corrections will be henceforth omitted.

For the Hamiltonian in Eq.~(\ref{eq7}) the matrix elements of the operators on the right-hand side of Eq.~(\ref{eq9}) are evaluated using
\begin{align}
\langle p| 1|x_i\rangle&=\langle p|x_i\rangle\\
\langle p|\hat{p}^2|x_i\rangle&=\langle p|\hat{p}^2|p\rangle \langle p|x_i\rangle=p^2\langle p|x_i\rangle\\
\langle p|V(\hat{x})|x_i\rangle&=\langle p|x_i\rangle\langle x_i| V(\hat{x})|x_i\rangle=V(x_i)\langle p|x_i\rangle\, .
\end{align}
The short-time propagator in Eq.~(\ref{eq9}) can now be approximated as
\begin{align}
\langle p|e^{-i\hat{H}\delta t/\hbar}|x_i\rangle
& \approx\bigg[1-{ip^2\delta t\over 2m\hbar}-{i\delta t\over\hbar}V(x_i)\bigg]\langle p|x_i\rangle \\
&\approx\exp\bigg\{-{i\over\hbar}\bigg[{p^2\delta t\over2m}+V(x_i)\delta t\bigg]\bigg\}\langle p|x_i\rangle\, ,
\end{align}
with the approximations becoming equalities for infinitesimal $\delta t$. We use 
\begin{equation}
\langle p|x\rangle={e^{-ipx/\hbar}\over\sqrt{2\pi\hbar}}\, ,
\label{eq14}
\end{equation}
to obtain
\begin{align}
\langle p|e^{-i\hat{H}\delta t/\hbar}|x_i\rangle 
={1\over\sqrt{2\pi\hbar}}\exp\bigg\{-{i\over\hbar}\bigg[px_i+{p^2\delta t\over2m}+V(x_i)\delta t\bigg]\bigg\}\, .
\end{align}
We return to the right-hand side of Eq.~(\ref{eq8}) and invoke Eq.~(\ref{eq14}) to find,
\begin{align}
\langle x_f,t_i+\delta t|x_i,t_i\rangle & =\!\int{dp\over2\pi\hbar}\exp\bigg\{-{i\delta t\over\hbar}\bigg[{p(x_i-x_f)\over\delta t}
+{p^2\over2m}+V(x_i)\bigg]\bigg\} \\
&=\sqrt{m\over2\pi i\hbar\delta t}\exp\bigg\{{i\over\hbar}\bigg[{m(x_f-x_i)^2\over2\delta t}-V(x_i)\delta t\bigg]\bigg\}\, .
\label{eq16}
\end{align}
The integral has been evaluated by completing the square in the argument of the exponential. If we make the identification
\begin{equation}
\bigg({dx\over dt}\bigg)^2=\bigg({x_f-x_i\over\delta t}\bigg)^2\, ,
\end{equation}
we see that the argument of the exponential on the right-hand side of Eq.~\eqref{eq16} is the product of $\delta t$ and the classical Lagrangian $L$:
\begin{equation}
L\,\delta t=\bigg [\frac{m}{2}\bigg({x_f-x_i\over\delta t}\bigg)^2-V(x_i)\bigg]\delta t\, .
\end{equation}
Hence, the short-time propagator reduces to
\begin{equation}
\langle x_f,t_i+\delta t |x_i,t_i\rangle=\sqrt{m\over2\pi i\hbar\delta t}\,e^{iL\delta t/\hbar}\, .
\label{eq19}
\end{equation}

We can now evaluate propagators over finite times by dividing the time interval into slices of duration $\delta t$,
\begin{align}
&\langle x_f,t_f|x_i,t_i\rangle=\!\iint\cdots \!\int\langle x_f,t_f|x_{N-1},t_{N-1}\rangle \nonumber \\
& \quad\times\langle x_{N-1},t_{N-1}|x_{N-2},t_{N-2}\rangle\cdots\langle x_2,t_2|x_1,t_1\rangle\nonumber\\
& \quad \times\langle x_1,t_1|x_i,t_i\rangle\,dx_1\,dx_2\cdots dx_{N-1}\, ,
\end{align}
and applying Eq.~(\ref{eq19}) to each slice:
\begin{equation}
\langle x_f,t_f|x_i,t_i\rangle=\!\int\prod_{n=1}^{N-1}dx_n\exp\bigg[{i\delta t\over\hbar}\sum_{n=1}^{N-1}\lcal(t_n)\bigg]\, . \label{thisexpression}
\end{equation}
We have omitted the prefactors in Eq.~\eqref{thisexpression} because they will not be needed in the following.

In the continuum limit ($N \to \infty$, $\delta t\to 0$, such that the product $N\delta t$ is fixed), the integral over positions at each time is the same as the integral over all paths between the initial and final positions:
\begin{equation}
\langle x_f,t_f|x_i,t_i\rangle=\!\int \dd x(t)\,e^{-i\s/\hbar}\, ,
\label{eq22}
\end{equation}
where $\dd x(t)\equiv\prod_{n=1}^{N-1} dx_i$ and, as $N\to\infty$, the action $S$ of the path $x(t)$ becomes
\begin{align}
S &=\!\int_{t_i}^{t_f}\lcal(x(t))\,dt
 =\!\int_{t_i}^{t_f}\bigg[{m\over2}\bigg({dx\over dt}\bigg)^2-V(x(t))\bigg]\,dt\, .
\label{eq23}
\end{align}

\subsection{Imaginary time path integrals}

The path integral in Eqs.~(\ref{eq22}) and (\ref{eq23}) yields transition amplitudes as the sum of the phases of all paths between the given initial and final positions. For our purposes {\it imaginary time} path integrals, where the time $t$ is replaced by $-i\tau$, with $\tau$ real, are of primary interest.

There are two main applications of imaginary time path integrals. In statistical mechanics $\tau=\hbar/(k_B T)$, where $k_B$ is Boltzmann's constant and $T$ is the absolute temperature. Thus, for equal initial and final positions $x$, an integration over $x$ produces the partition function $Z$:
\begin{equation}
Z=\!\int\big\langle x\big|e^{-\hat{H}\tau/\hbar}\big|x\big\rangle\,dx=\mbox{Tr}\big(e^{-\hat{H}\tau/\hbar}\big)\, ,
\end{equation}
in which the trace Tr is the sum/integral of the diagonal elements of an operator.

Another application is the determination of the energy spectrum of a quantum system. This calculation utilizes the identity $1=\sum_n|n\rangle\langle n|$ in terms of the eigenfunctions of the Hamiltonian, such that $\hat{H}|n\rangle=E_n|n\rangle$,
\begin{align}
\z&=\!\int\big\langle x \big|e^{-\hat{H}\tau/\hbar}\big|x\big\rangle\,dx =\!\int\sum_n\big\langle x\big| e^{-\hat{H}\tau/\hbar}\big|n\big\rangle\big\langle n\big|x\big\rangle\,dx  \\
&=\!\int\sum_n e^{-E_n \tau/\hbar}\psi_n(x)\bar{\psi}_n(x) dx =\sum_n e^{-E_n\tau/\hbar},
\end{align}
where we have used the fact that $\psi_n(x)$ is normalized. Similarly, we can expand the propagator $\langle x_f, t_f | x_i, t_i\rangle$ in terms of the eigenfunctions $\{|n\rangle\}$:
\begin{equation}\label{expProp}
\langle x_f, t_f | x_i, t_i\rangle = \sum_{n=0}^{\infty} e^{-E_n (t_f - t_i)/\hbar} \langle x_f | n\rangle \langle n | x_i\rangle.
\end{equation}

The derivation of the imaginary-time path integral proceeds along the same lines as the real-time propagator, with the result corresponding to Eq.~(\ref{eq16}) given by
\begin{align}
&\langle x_f| e^{-\hat{H}\delta\tau/\hbar}|x_i\rangle 
=\sqrt{m\over2\pi\hbar\delta\tau}\exp\bigg\{-{\delta\tau\over\hbar}\bigg[{m\over2}\bigg({x_f-x_i\over\delta\tau}\bigg)^2+V(x_i)\bigg]\bigg\}\, , \label{thisresult}
\end{align}
In the limit $N\to\infty$, Eq.~\eqref{thisresult} can be used to write the partition function in a form analogous to Eqs.~(\ref{eq22}) and (\ref{eq23}):
\begin{equation}
Z=\mbox{Tr} \big( e^{-\hat{H}(\tau_f-\tau_i)/\hbar}\big)=\!\int \dd x(\tau)\,e^{-\s/\hbar}\,,
\label{eq28}
\end{equation}
where $\s$ is the (Euclidean) action over a path $x(\tau)$ with $\tau_f\geq\tau\geq\tau_i$, and $x(\tau_f)=x_f,~x(\tau_i)=x_i$.
\begin{align}
S=\!\int_{\tau_i}^{\tau_f}\lcal(x(\tau)) \,d\tau  =\!\int_{\tau_i}^{\tau_f}\bigg[{m\over2}\bigg({dx\over d\tau}\bigg)^2+V(x(\tau))\bigg]\,d\tau\, .
\label{eq29}
\end{align}
The integrals in Eqs.~(\ref{eq28}) and (\ref{eq29}) and their real-time counterparts in Eqs.~(\ref{eq22}) and (\ref{eq23}) are over all paths weighted by Lagrangian-type quantities. However, in the imaginary-time formalism, quantities associated with the paths are real. 

\subsection{The quantum harmonic oscillator}

The Hamiltonian for a particle of mass $m$ bound by a harmonic potential with force constant $k$ is
\begin{equation}
\hat{H}={\hat{p}^2\over2m}+{k\hat{x}^2\over2}={\hat{p}^2\over2m}+{m\omega^2\hat{x}^2\over2}\, ,
\end{equation}
where $\omega=\sqrt{k/m}$ is the natural frequency of the oscillator. The discretized Euclidean Lagrangian for this system is
\begin{equation}
\lcal_i={m\over2}\bigg({x_{i+1}-x_i\over\delta\tau}\bigg)^2+{m\omega^2x_i^2\over2}\, ,
\end{equation}
which allows us to express the Euclidean action and the partition function as
\begin{align}
S &=\sum_{n=1}^{N-1}\lcal_i \\
Z&=\!\int_{-\infty}^\infty\prod_{i=1}^{N-1}dx(\tau_i)\exp\bigg(-{\delta\tau\over\hbar}S\bigg)\, .
\label{def_s}
\end{align}
The energy eigenvalues of $\hat{H}$ are $E_n=\hbar\omega (n+\frac{1}{2})$ for $n=0, 1, 2,\ldots$\,. The normalized ground state wave function is
\begin{equation}\label{wavefunction}
\psi_0=\left(\frac{m\omega}{\pi\hbar}\right)^{\frac{1}{4}}\exp\left(-\frac{m\omega x^2}{2\hbar}\right),
\end{equation}
from which all other wave functions can be obtained through ladder operations. Expectations of observable quantities in the ground state are determined by $\psi_0(x)$.

\section{\label{sec3}Computational method}

The formalism discussed in Sec.~\ref{sec2} will be applied to the harmonic oscillator. However, the range of applicability is much broader. The idea is that if the partition function can be constructed (``if the system can be simulated''), an arbitrary observable can be determined (``measured'') with a statistical uncertainty that decreases as the simulation is extended. For the construction of such observables and their evaluation the complete tool set of statistical mechanics can be used. An overview of all parameters and their meanings is given in Table~\ref{tableA}.

\begin{table}[t]
 \begin{ruledtabular}
\begin{tabular}{  l l }
 \multicolumn{1}{c}{Parameter} & \multicolumn{1}{l}{Meaning} \\ \hline
$ N_{\tau}$  & number of elements of the time lattice \\ 
$i\delta\tau$ & Euclidean time, with $i\in\{1,\ldots,N_\tau\}$ the \\
~ & \quad site index \\
$t_{\rm MC}$ & Monte Carlo time; refers to index of a \\
~ & \quad path in the Markov chain \\
sweep & $N_\tau$ applications of the \\ 
~ & \quad single-site Metropolis--Hastings algorithm  \\
$N_{\mathrm{sep}}-1$ & number of discarded paths between\\ 
~ & \quad successive paths used for measurement \\
$\delta\tau$ & lattice spacing \\
$\tilde{m}$ & dimensionless effective mass: $\tilde{m}=m\delta\tau$ \\
$\tilde{\omega}$ & dimensionless frequency: $\tilde{\omega}=\omega\delta\tau$\\
$N$ & number of paths within an ensemble\\
$\xi$ & ``correlation time'': Euclidean time for \\
~ & \quad two-point correlations to diminish by \\
~ & \quad a factor $e$ \\
$m_{\rm eff}$ & effective mass: $m_{\rm eff}=1/\xi$ \\
$N_B$ & number of bins in jackknife procedure \\
$B$ & bin width $B=N/N_B$ \\
$\tau_{O, \exp/{\rm int}}$ & exponential and integrated \\
~& \quad autocorrelation time of  observable $O$
\end{tabular}
\end{ruledtabular}
\caption{\label{tableA} An overview of all parameters and their meanings.}
\end{table}

\subsection{Monte Carlo methods}
\begin{figure}[t]
\centering
\includegraphics[width=7cm]{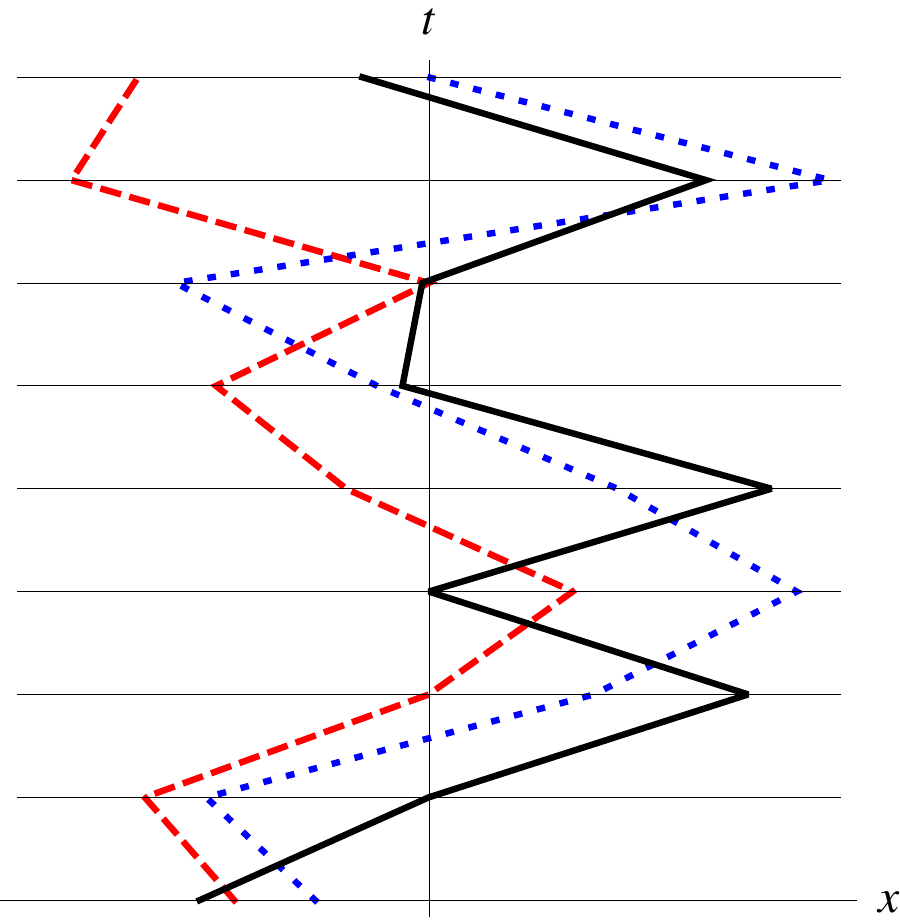}
\caption{(Color online) Illustration of the computational method. The spatial location of the particle at time $\tau_i$, where $i=1,\ldots,N_{\tau} = 8$, can take any real value,  but is constrained by the potential centered at the origin and its neighboring positions at times $\tau_{i-1}$ and $\tau_{i+1}$. The solid line represents the thermalized path$^{(0)}$; the dotted line is the next path in the Markov chain, path$^{(1)}$; and the dashed line is the resulting trajectory after 19 further Metropolis sweeps.}
\label{diagram}
\end{figure}

The simulation is done on a discrete time lattice with $N_{\tau}$ time slices with periodic boundary conditions so that the time slice $N_{\tau}+1$ equals the time slice $1$. To calculate the statistics of the observables, many particle trajectories of the form $(x_1,\ldots,x_{N_{\tau}})$ are needed, where each coordinate is a real number. Starting from an initial, thermalized configuration path$^{(0)}$ (see Sec.~\ref{thermalization}), the path is updated by the Metropolis--Hastings algorithm. The application of this elementary update to the variable $x_i$ for each time slice $i$ constitutes one ``sweep'' or one Monte Carlo step per site. One Metropolis sweep yields the next path in the sequence, e.g., path$^{(1)}$ from path$^{(0)}$. Because path$^{(\nu)}$ relies only on path$^{(\nu-1)}$, the trajectories constitute a Markov chain. The computational method is illustrated in Fig.~\ref{diagram}. We distinguish between Euclidean time $\tau=i\delta\tau$, which indicates the index $i$ of a lattice site, and Monte Carlo time, which refers to the index of a path $\nu$ in the Markov chain. Because each path in the chain is based on the previous path, the paths are correlated. We will discuss these correlations in more detail in Sec.~V. To combat the autocorrelation, we discard a number $N_{\mathrm{sep}}-1$ of paths between every two paths used for measurements. We will refer to the remaining paths used to calculate average quantities as ``configurations.''

\subsection{\label{Dimensionless} Dimensionless variables and observables}
Because computer code can handle only pure numbers, it is necessary to express the physics of the system in dimensionless form. A naive way of doing this is by expressing all quantities in metric units, e.g., meters. A  disadvantage of this choice is that it leads to numbers that often span several orders of magnitude. To avoid this problem, we  express all variables in terms of the appropriate power of the lattice spacing $\delta\tau$. 
To this end, we set $\hbar=1=c$, which implies that
$[\mathrm{time}]=[\mathrm{length}]=[\mathrm{mass}^{-1}]=[\mathrm{energy}^{-1}]$.
We introduce the dimensionless variables:
\begin{equation}
\tilde{m}=m\delta \tau, \quad \tilde{\omega}=\omega\delta \tau, \quad \tilde{x}_i=\frac{x_i}{\delta \tau},
\end{equation}
The dimensionless action becomes
\begin{equation}
\label{rescaling}
{\tilde S} =\sum_{i=1}^{N_{\tau}}\left[ \frac{1}{2}\tilde{m}({\tilde{x}}_{i+1}-{\tilde{x}}_i)^2+\frac{1}{2}\tilde{m}\tilde{\omega}^2\tilde{x}_i^2 \right],
\end{equation}
where $\tilde{m}$, $\tilde{\omega}$ and $\{\tilde{x}_i\}$ are dimensionless.
Note that the summation range differs from the one in Eq.~(\ref{def_s}), due to  periodic boundary conditions.
We restrict ourselves to the subspace $\tilde{m}=\tilde{\omega}$ of the available parameter space. The parameter $\tilde{m}$ can thus be viewed as the effective lattice spacing of a harmonic oscillator with unit mass and unit natural frequency.

The continuum limit applies to observables and is not taken for an individual simulation. Rather, we run a series of simulations, each with a smaller effective lattice spacing (that is, a smaller value of $\tilde{m})$ and a greater value of $N_{\tau}$ than the previous, such that the product $N_{\tau}\tilde{m}$ is fixed.

For a quadratic action, all odd moments of $\hat{x}$ have zero expectation value. An analytic expression for $\langle\hat{x}^2\rangle$ is derived in Ref.~\onlinecite{Creutz}:
\begin{equation}\label{x2}
\langle\hat{x}^2\rangle=\frac{1}{2\tilde{m}\tilde{\omega}\sqrt{1+{1\over4}\tilde{\omega}^2}}\left(\frac{1+R^{N_{\tau}}}{1-R^{N_{\tau}}}\right),
\end{equation}
with the auxiliary variable
\begin{equation}\label{aux}
R=1+\frac{\tilde{\omega}^2}{2}-\tilde{\omega}\sqrt{1+{\tilde{\omega}^2\over4}}.
\end{equation}
The other observable we need is
\begin{equation}\label{x4}
\langle\hat{x}^4\rangle=\frac{3}{(2\tilde{m}\tilde{\omega})^2(1+{1\over4}\tilde{\omega}^2)}\left(\frac{1+R^{N_{\tau}}}{1-R^{N_{\tau}}}\right)^2=3\langle\hat{x}^2\rangle^2.
\end{equation}
Derivations of the expressions for Eqs.~(\ref{x2}) and (\ref{x4}) are given in Ref.~\onlinecite{SI}.

\subsection{\label{Metropolis}The Metropolis update}

The core of our path-generating algorithm is an update of a single site based on the Metropolis--Hastings algorithm.\cite{Metropolis, hastings70} The output is a set of $N$ paths $\{\tilde{x}_1,\ldots,\tilde{x}_{N_{\tau}}\}$ with Boltzmann weights $\rho[\{\tilde{x}_i\}]\sim\exp[-{\tilde S}[\{\tilde{x}_i\}]]$. The input of the Metropolis update is an array path with $N_{\tau}$ sites, a real number $h$, and the parameters $\tilde{m}$ and $\tilde{\omega}$. Periodic boundary conditions avoid the need to abandon (in the data)  sites   affected by the lattice edges. One sweep visits $N_{\tau}$ sites in random order. A site may be visited repeatedly or not at all, but the mean number of visits per sweep for each site is one. Each Metropolis update to a given site consists of four steps.
\begin{enumerate}
\item Generate a random number $u$ from a uniform distribution in the interval $[-h,h]$.
\item Propose a change to the visited site, $\tilde{x}_i \to \tilde{x}_i'=\tilde{x}_i+u$.
\item Compute the change in the action $\delta {\tilde S}$ as a result of this trial modification.
\item Accept the change with probability $\min\{1, e^{-\delta {\tilde S}}\}$.
\end{enumerate}
Pseudocode is provided in Ref.~\onlinecite{SI}.
The probability $\min\{1, e^{-\delta {\tilde S}}\}$ in step 4 implies that proposed modifications that lower the action are always accepted. A trial that would increase the action is accepted with probability $e^{-\delta {\tilde S}}$. This decision is made in an accept/reject step. The Metropolis update satisfies detailed balance:\cite{Morningstar}
\begin{equation}
p(\tilde{x}_i\to \tilde{x}_i^\prime) e^{-{\tilde S}(\tilde{x}_i)}=p(\tilde{x}_i^\prime\to \tilde{x}_i) e^{-{\tilde S}(\tilde{x}_i')}\, .
\end{equation}
Because of this property, the estimated average of an observable $\hat{O}$ reduces to an arithmetic average. After one sweep, the acceptance ratio is computed. The value of $h$ is adjusted to meet a predefined acceptance ratio. We chose the desired acceptance ratio to be 0.8 (which is a conventional choice in lattice QCD), even though we suspect the ideal value for the harmonic oscillator, with the choices of $\tilde{m}=\tilde{\omega}$ listed in Table~II, to be  smaller.  Although algorithms with too low or too high an acceptance ratio are less efficient, the generated Boltzmann distribution of paths is unaffected by this choice.

\subsection{Thermalization}
\label{thermalization}

\begin{figure}[t]
\centering
\includegraphics[width=7cm]{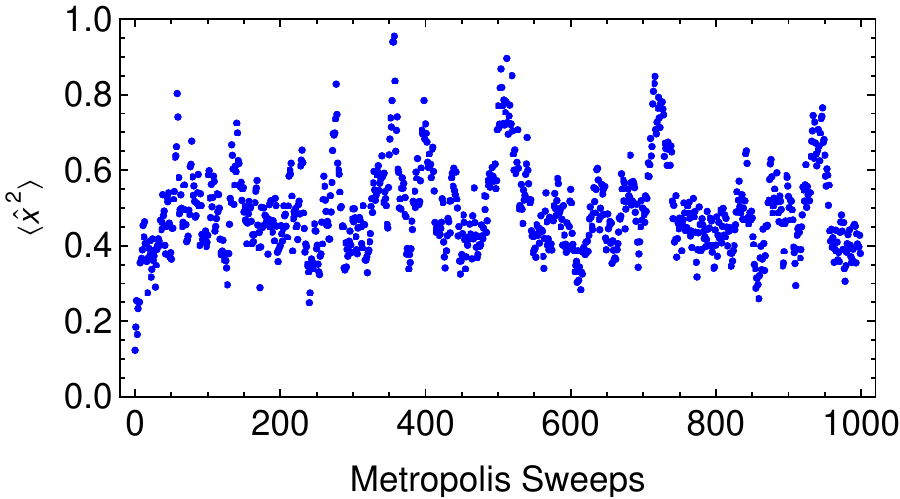}
\caption{(Color online) A trial run for $\langle\hat{x}^2\rangle$ to illustrate thermalization effects ($\tilde{m}=\tilde{\omega}=0.1$, $N_{\tau}= 1200$). One thousand paths were discarded between every two configurations whose output is shown. The first 50--100 configurations should not be used for measurements.}
\label{therm}
\end{figure}

The required thermalization process can start from an array of zeros (a ``cold'' start), random numbers (a ``hot'' start), or an initial path that is expected to be close to a thermalized path. The initial thermalization steps are not characteristic of the probability density $\rho[\{\tilde{x}_i\}]\sim \exp[-{\tilde S}\{\tilde{x}_i\}]$ and must be discarded lest they skew the simulation. A trial run is one way to choose the number of sweeps needed before the array qualifies as a thermalized path. An example of a thermalization process is given in Fig.~\ref{therm}. For each configuration, the observable $\langle\hat{x}^2\rangle$ was measured to monitor its fluctuations around the expected values.\cite{f2} In our code we first average $\tilde{x}_i^2$ over the $N_{\tau}$ time slices in a given configuration, and then   compute the ensemble average of that number over the configurations. For actions for which the exact answer is not known, independent ``hot'' and ``cold'' runs can help to establish the expectation value. In this case, the first 50--100 configurations should not be used.

\subsection{Two-point correlation function}\label{sec:twopoint}
\begin{figure}[t]
\includegraphics[width=7cm]{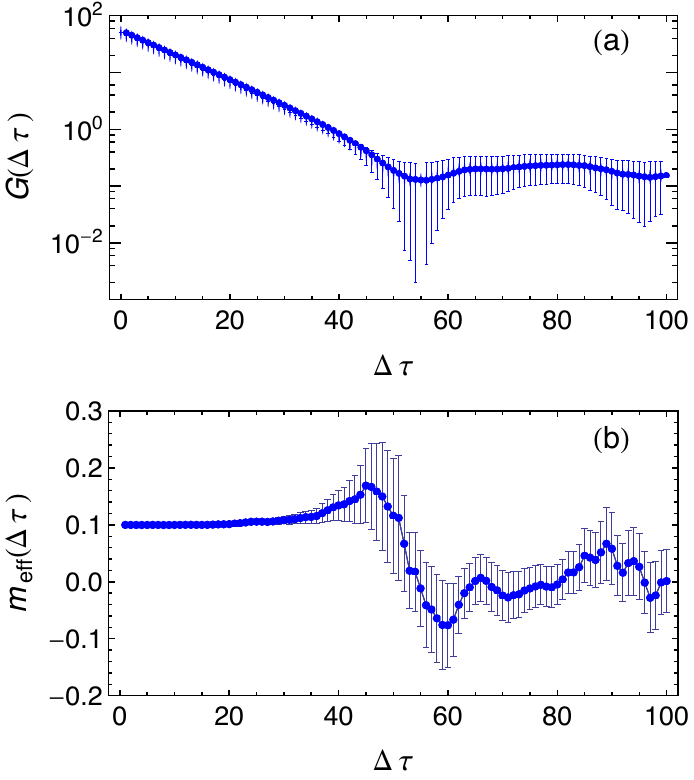}
\caption{(a) The symmetrized two-point correlation function ($N_{\tau}=1200$, $\tilde{m}=\tilde{\omega}=0.1$, $N=10^4$). The exponential decrease is swamped by noise after approximately 40 time slices and the magnitude of the error bars starts to increase significantly, and eventually the error bars become unreliable. (b) The effective mass $1/\tilde{\xi}$. The estimates of the errors  are reliable until $\Delta \tau\approx 40$.\cite{f4}}
\label{fig2}
\end{figure}

To make optimal use of CPU time, it is important to choose the number of sites $N_{\tau}$ as small as possible, but large enough to avoid finite-size effects. How is the lower bound on $N_{\tau}$ established? Correlations within a lattice are quantified by the connected two-point function:
\begin{equation}
G(\Delta \tau)=\braket{x(\tau)x(\tau+\Delta \tau)}-\braket{x(\tau)}\braket{x(\tau+\Delta \tau)},\,
\end{equation}
where we have written $x(\tau)$ instead of $x_i$ to emphasize the  dependence of $G$ on the time difference $\Delta \tau$, where $\tau$ and $\Delta\tau$ can be any multiple of the lattice spacing. Because $\braket{x(\tau)}=0$ for all $\tau$ for the harmonic oscillator, we work with the two-point function
\begin{align}\label{normaltwopt}
G(\Delta \tau)&=\braket{x(\tau) x(\tau+\Delta \tau)}\\
&=\frac{1}{N_{\tau}}\sum_{i=1}^{N_{\tau}} \sum_{\substack{j \\ (j-i)\,\mathrm{mod}\,~N_{\tau}=\Delta \tau}} x(i) x(j).
\end{align}
An example of the exponential decay of $G(\Delta \tau)$ is shown in Fig.~\ref{fig2}(a),\cite{Gattringer}
\begin{equation}
\label{twoptexp}
G(\Delta \tau)=A e^{-\Delta \tau/\xi}\ + A e^{-(T-\Delta \tau)/\xi},
\end{equation}
where $\xi$ is the correlation time and $T$ is the final time. The second term in Eq.~(\ref{twoptexp}) is due to  periodic boundary conditions. The total length of the lattice must be greater than $\xi$. We choose $N_{\tau}$ to be about $10 \tilde{\xi}$ (where $\tilde{\xi}$ is the correlation time expressed in lattice units).

An estimate of $1/\tilde{\xi}$ can be obtained from the local logarithmic slope for suitable $\Delta \tau$:\cite{Gattringer, f3}
\begin{equation}
\frac{1}{\tilde{\xi}} = \frac{1}{2}\log\left[\frac{G(\Delta \tau-1)}{G(\Delta \tau+1)}\right]\, .
\end{equation}
The quantity $1/\xi$ is known as the effective mass $m_{\rm eff}$. Figure~\ref{fig2}(b) shows $\tilde{m}_{\rm eff}$ for the same set of paths used for Fig.~\ref{fig2}(a).

\begin{figure}[h]
\centering
\includegraphics[width=7cm]{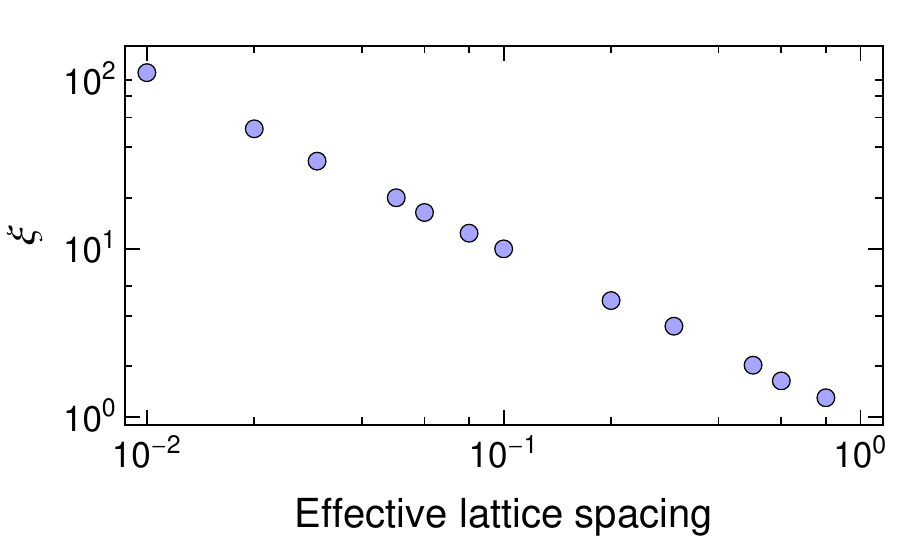}
\caption{\label{fig4} The correlation length $\xi$ \emph{versus} the effective lattice spacing ($N_{\mathrm{sep}}=300$, $N=10^4$). The points fall on a straight line, indicating a power-law dependence. The error bars were constructed using a jackknife analysis. Error bars are smaller than the symbol size.}
\end{figure}

Figure~\ref{fig4} suggests that there is a power-law dependence of $\tilde{\xi}$ on the effective lattice spacing. To construct Fig.~\ref{fig4}, we repeated the procedure illustrated in Fig.~3 for 13 effective lattice spacings listed in Table~\ref{table1}. The physical length, the product of $\tilde{m}=\tilde{\omega}$ and $N_{\tau}$, was kept constant. With this choice of parameters, we were able to explore two orders of magnitude in the lattice spacing; the associated $N_{\tau}$ are round numbers.

\begin{table}[t]
\centering
 \begin{tabular}{ | d | c || d | c |}
 \hline
 \multicolumn{1}{|c}{$\tilde{m}=\tilde{\omega}$} & \multicolumn{1}{|c||}{$N_{\tau}$} & \multicolumn{1}{c|}{$\tilde{m}=\tilde{\omega}$} & \multicolumn{1}{c|}{$N_{\tau}$}\\ \hline\hline
 1 & 120 & 0.1 & 1200 \\
 0.8 & 150 & 0.08 & 1500 \\
 0.6 & 200 & 0.06 & 2000 \\
 0.5 & 240 & 0.05 & 2400 \\
 0.3 & 400 & 0.03 & 4000 \\
 0.2 & 600 & 0.02 & 6000 \\
 & & 0.01 & 12000 \\
\hline
\end{tabular}
\caption{\label{table1} Effective lattice spacings used for the results shown in Figs.~4--8.}
\end{table}

\section{Jackknife analysis}\label{Jackknife}
\label{sec4}
Suppose we compute values $O_1,\ldots,O_N$, of an observable $\hat{O}$ (typically a moment of $\hat{x}$), with the expectation value and variance:
\begin{align}
\big<\hat{O}\big> &=\langle O \rangle\, \\
\big< \big(\hat{O} -\big<\hat{O}\big>\big)^2\big>&=\sigma_O^2\, .
\end{align}
The quantity $E(O)$ provides an unbiased estimator of the mean:
\begin{equation}\label{unbiasedestimator}
\E(O)=\E_N(O)=\frac{1}{N}\sum_{i=1}^N O_i\, ,
\end{equation}
where $N$ is the number of measurements. An unbiased estimator of the sample variance is given by
\begin{equation}\label{sigmastd}
\sigma^2_{O,\,\mathrm{std}}=\frac{1}{N-1}\sum_{i=1}^N [O_i-\E(O)]^2.
\end{equation}
This estimator is applicable even if the measurements are somewhat correlated, that is, not entirely independent of each other. The square root of the variance of the estimator of the sample mean is the error in our estimate, not the square root of the variance of the distribution. For the former quantity it matters whether the measurements are independent or not. For uncorrelated (independent) measurements the relation is
\begin{equation}
\label{jackvar}
\sigma^2_{\E(O),\,\mathrm{naive}}=\frac{\sigma_{O,\,\mathrm{std}}^2}{N}.
\end{equation}
The subscript ``naive'' refers to the assumption that the variables are not correlated. This assumption implies that
the statistical error of the mean in Eq.~(\ref{unbiasedestimator}) is given by 
\begin{equation}\label{errO}
{\rm err}_O=\sqrt{\sigma^2_{\E(O),\,\mathrm{naive}}}=\frac{\sigma_{O,\mathrm{std}}}{\sqrt{N}}.
\end{equation}

\begin{figure}[t]
\includegraphics[width=7cm]{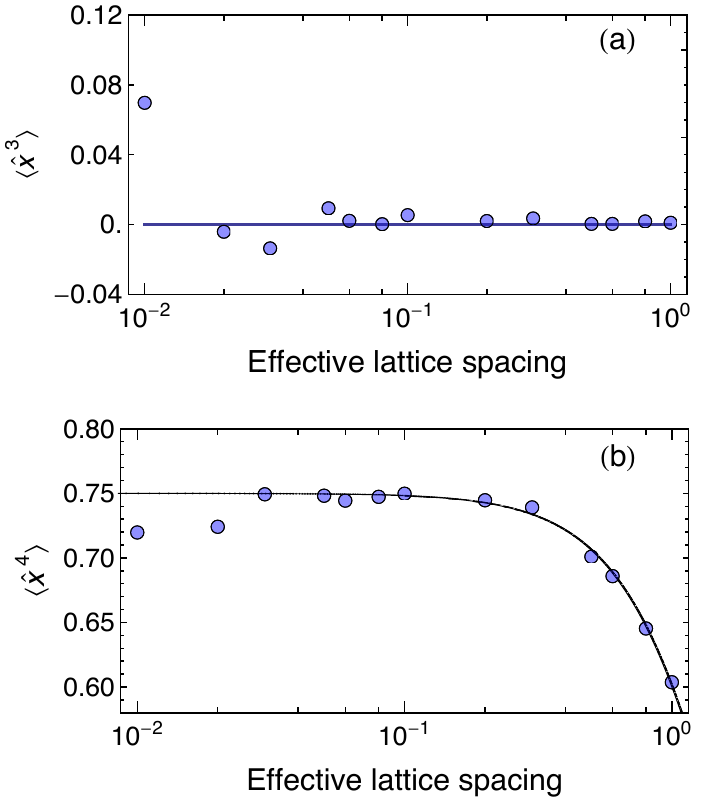}
\caption{(a) Simulations of $\langle\hat{x}^3\rangle$ as a function of the effective lattice spacing ($N_{\mathrm{sep}}=300$).  (b) Simulations of $\langle\hat{x}^4\rangle$ as a function of the effective lattice spacing (with the same parameters). The solid line is the exact result Eq.~(\ref{x4}). The error bars are smaller than the size of the symbols, because
the naive error ignores correlations between measurements.}
\label{fig3}
\end{figure}

For correlated data the factor of $N$ in the denominator of Eq.~(\ref{jackvar}) is replaced by $N_\mathrm{eff}<N$, and the statistical error of the mean differs from the naive estimate in Eqs.~(\ref{jackvar}) and \eqref{errO}. That is, for correlated data the naive estimate underestimates the true statistical error of the sample mean as shown in Fig.~\ref{fig3} for $\braket{\hat{x}^3}$ and $\braket{\hat{x}^4}$. The statistical error is smaller than the size of the dots in Fig.~\ref{fig3}, yet many of the dots do not lie on the exact curve. For a fixed number of sweeps between adjacent measurements ($N_{\mathrm{sep}}=300$) the problem is seen to worsen for smaller effective lattice spacings.

The jackknife procedure\cite{Quenouille} provides a more realistic estimate of the variance of the mean of a set of correlated variables. The jackknife variance of a parameter is found by systematically leaving out batches of observations from a dataset, calculating the variance each time a different batch is omitted, and  finding the average of these variance calculations. A theoretical justification of this procedure is given in Ref.~\onlinecite{Shao}.  The $N$ samples are divided into $N_B$ blocks of bin width $B$. The block estimators are
\begin{equation}\label{JackBlockEst}
o_k=\frac{1}{B}\sum_{i=1}^B O_{(k-1)B+i} \quad (k=1,\ldots N_B).
\end{equation}
The bin width $B$ should exceed the autocorrelation time of the observable to ensure that the $N_B$ values can be treated as uncorrelated.\cite{f6} The bin-based variance of the mean is given by
\begin{equation}
\label{binspec}
\sigma^2_{\E(O),\,\mathrm{bins}}=\frac{1}{N_B(N_B-1)}\sum_{k=1}^{N_B}[o_k-\E(O)]^2.
\end{equation}
The estimator $o_k$ in Eq.~(\ref{binspec}) is an average over only the $N_B\mathrm{th}$ fraction of all the measurements.
This limited number of measurements may prevent the determination of $o_k$ for some $k$ (for instance a fit based on too few configurations may occasionally fail to converge). \cite{Wolff}
This problem is overcome by using complementary bins:
\begin{equation}\label{compbins}
\tilde{o}_k=\frac{1}{N-B}\left(\sum_{i=1}^N O_i- Bo_k\right).
\end{equation}
Rather than $N_B$ estimators $o_1,\ldots, o_{N_B}$, each containing $B$ measurements, as in Eq.~(\ref{JackBlockEst}), we work with $N_B$ jackknife estimators $\tilde{o}_1,\ldots, \tilde{o}_{N_B}$, each based on $N-B\gg B$ measurements. The resulting complementary bin-based or jackknife variance of the mean is 
\begin{equation}\label{JackErrorEst}
\sigma^2_{\E(O),\mathrm{jack}}=\frac{N_B-1}{N_B}\sum_{k=1}^{N_B}[\tilde{o}_k-\E(O)]^2.
\end{equation}

\begin{figure}[t]
\includegraphics[width=7cm]{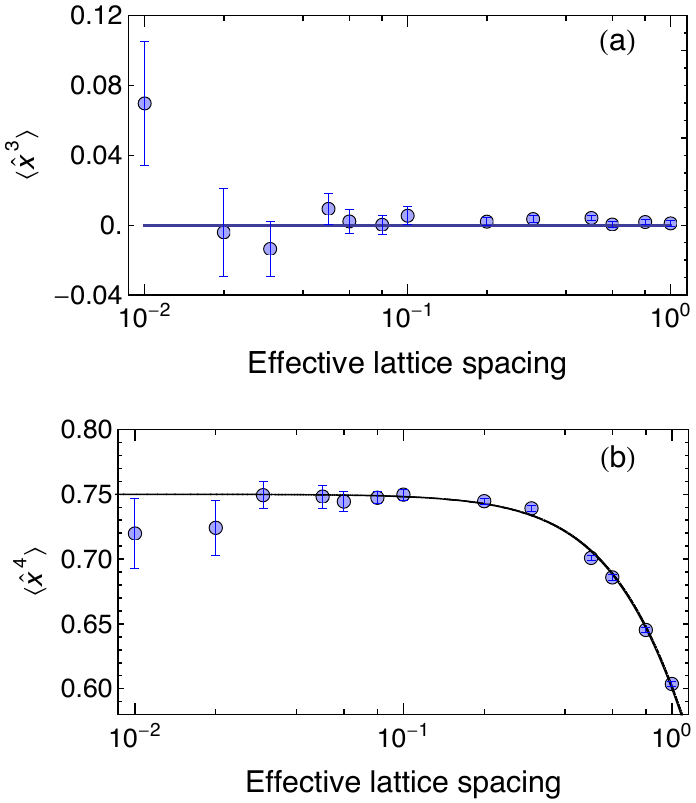}
\caption{(a) Simulations of $\langle\hat{x}^3\rangle$ as a function of the effective lattice spacing ($N_{\mathrm{sep}}=300$) with jackknife error bars, which account for correlations within the data. (b) Simulations of $\langle\hat{x}^4\rangle$ as a function of the effective lattice spacing ($N_{\mathrm{sep}}=300$) with jackknife error bars. The data are always within two jackknife errors of the theoretical results  in Eq.~(\ref{x4}).}
\label{fig6}
\end{figure}

Figure~\ref{fig6} features the same data set as Fig.~\ref{fig3}, but this time the statistical errors are determined as $\sigma_{\E(O),\,\mathrm{jack}}$ [see Eq.~(\ref{JackErrorEst})]. The errors are assessed reliably; the $\sigma_{\E(O),\,\mathrm{jack}}$ error bars miss the analytical curve at a frequency consistent with the expected $32\%$. This figure corresponds to the relative area outside the $\pm 1\sigma$ band around the center of a normal distribution.

Note that for $\langle\hat{x}^3\rangle$ and $\langle\hat{x}^4\rangle$ the bin-based estimator (\ref{binspec}) would have served the same purpose. 
The jackknife formula (\ref{JackErrorEst}) is built to reproduce the result of (\ref{binspec}) whenever an observable can be determined from a single configuration. 
In general, this is not true (e.g. for ``secondary'' observables derived from ``primary'' observables through a fit). 
It is, therefore, common practice to use the jackknife procedure by default.

\section{Autocorrelation time}
\label{Autocorrelation}

The correlation of a sequence of generated configurations arises naturally because one configuration differs from the next only by the result of a fixed number of sweeps. The autocorrelation time provides information on how strongly subsequent measurements are correlated. Because correlations lead to increased errors in measurements, its accurate assessment is important. The autocorrelation for an observable $\langle\hat{O}\rangle$, which takes values $\{O_i\}$, as a function of Monte Carlo time $t_{\mathrm{MC}}$ is defined as: 
\begin{align}
\label{autocorr}
A_O(t_{\mathrm{MC}}) 
&=\E[(O_i- \E(O_i))(O_{i+t_{\mathrm{MC}}}-\E(O_{i+t_{\mathrm{MC}}})]\\
&=\frac{1}{N-t_{\mathrm{MC}}-1}
\sum_{i=1}^{N-t_{\mathrm{MC}}}[O_i-\E(O_i)][O_{i+t_{\mathrm{MC}}}-\E(O_{i+t_{\mathrm{MC}}})] \, ,
\end{align}
where the average $\E(O_i)$ is over the first $N-t_{\mathrm{MC}}$ measurements and the average $\E(O_{i+t_{\mathrm{MC}}})$ over the last $N-t_{\mathrm{MC}}$ measurements. Note that $A_O(0)=\sigma_{O,\mathrm{std}}^2$.Comparisons between autocorrelation times for different observables are easier to make when the normalized $A_O(t_{\mathrm{MC}})/A_O(0)$ is considered instead.

Two parameters can be extracted from Eq.~(\ref{autocorr}): the asymptotic (or exponential) and the integrated autocorrelation times. The autocorrelation function of $O$ typically exhibits multi-exponential behavior:\cite{Gattringer}
\begin{equation}
\label{autocorrseries}
\frac{A_O(t_{\mathrm{MC}})}{A_O(0)}=a_0 e^{-t_{\mathrm{MC}}/\tau_0}+a_1 e^{-t_{\mathrm{MC}}/\tau_1}+\cdots\, ,
\end{equation}
with $\tau_0<\tau_1<\tau_2<\cdots$.
Usually $a_0\gg a_1\gg a_2\gg \cdots$,
where $a_0+a_1+a_2+\cdots=1$. Determining the ``true'' exponential autocorrelation time, defined as $\max(\tau_0, \tau_1, \tau_2,\cdots)$, requires precise data at large $t_{\mathrm{MC}}$, which are not normally available. To obtain an estimate for the exponential autocorrelation time, we make a multi-exponential fit as in Eq.~(\ref{autocorrseries}). In practice, such a fit is likely to contain one or two terms.

To determine the integrated autocorrelation time, we start from the variance of the unbiased estimator of the mean. Let $t_{\mathrm{MC}}$ be the absolute time difference between measurement $i$ and $j$, such that
\begin{align}
\label{tauint}
\left( \frac{1}{N}\sum_{i=1}^N O_i-\E(O_i) \right)^2
&=\frac{1}{N^2}\sum_{i=1}^N\sum_{j=1}^N [O_i-\E(O_i)][O_j-\E(O_j)] \\
&=\frac{1}{N}\sigma^2_{O,\,\mathrm{std}}
+\frac{1}{N^2}\sum_{i=1}^N\sum_{j\neq i} [O_i-\E(O_i)](O_j-\E(O_j)] \\
&=\frac{1}{N}\sigma^2_{O,\,\mathrm{std}}+\frac{2}{N^2}\sum_{i=1}^{N-1}\sum_{j=i+1}^N [O_i-\E(O_i)] [O_j-\E(O_j)] \\
&=\frac{\sigma_{O,\,\mathrm{std}}^2}{N}\Big\{1+ \frac{2}{N}\frac{1}{\sigma^2_{O,\,\mathrm{std}}}
\sum_{t_{\mathrm{MC}}=1}^{N-1}\sum_{i=1}^{N-t_{\mathrm{MC}}} [O_i-\E(O_i)][O_{i+t_{\mathrm{MC}}}-\E(O_{i+t_{\mathrm{MC}}})]\Big\} \nonumber\\
&=\frac{2\sigma_{O,\,\mathrm{std}}^2}{N}\left\{\frac{1}{2}+\frac{N-t_{\mathrm{MC}}}{N}\sum_{t_{\mathrm{MC}}=1}^{N-1}\frac{A_O(t_{\mathrm{MC}})}{A_O(0)}\right\}.
\end{align}
For $N\gg 1$ the right-hand side of Eq.~(\ref{tauint}) approaches$(2\sigma_{O,\mathrm{std}}^2/N)\tau_{O,\,\mathrm{int}}$,
where
\begin{equation}
\tau_{O,\,\mathrm{int}}=\frac{1}{2}+\sum_{t_{\mathrm{MC}}=1}^{N-1}\frac{A_O(t_{\mathrm{MC}})}{A_O(0)}.
\end{equation}
Comparison of Eq.~(\ref{tauint}) with the naive variance of the mean in Eq.~(\ref{jackvar}) shows that the effective number of independent measurements is $N_{\mathrm{eff}}=N/(2\tau_{O,\,\mathrm{int}})$.

To compute the integrated correlation time, we have to cut off the sum where the exponential relation for $A_O(t_{\mathrm{MC}})$ breaks down. If the cutoff is clearly defined (for example, at the first value where the autocorrelation becomes negative), the integrated autocorrelation time takes a unique value for a given data set, and is thus less subjective than the exponential correlation time. However, the latter clearly captures the exponential behavior of the autocorrelations.

\begin{figure}[t]
\includegraphics[width=7cm]{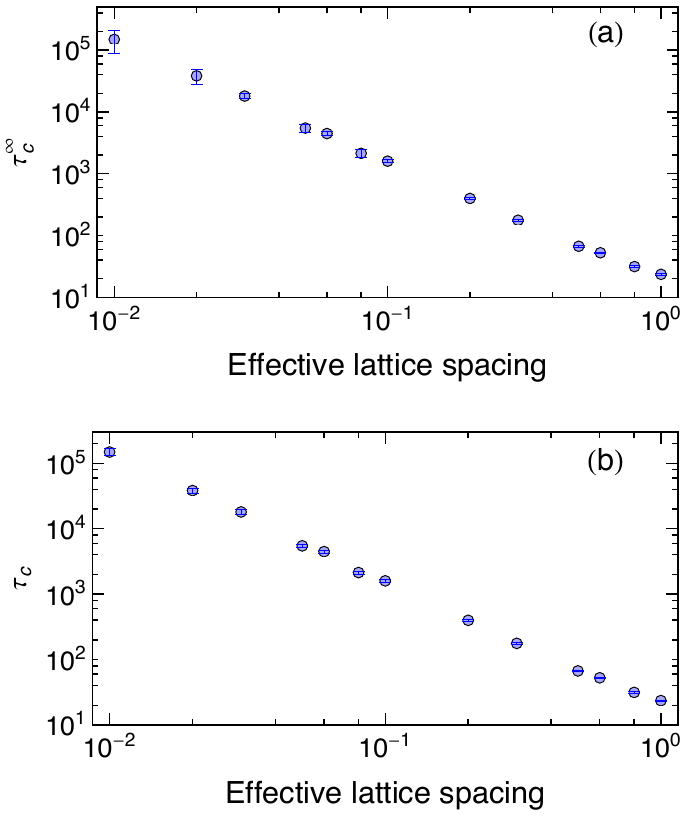}
\caption{(a) Number of lattice updates required to generate a path independent of its predecessor as a function of the lattice spacing $\tilde{m}$. The exponential autocorrelation time\cite{UWerr} of the observable $\hat{x}$ exhibits power-law behavior as a function of $\tilde{m}$, $\tau_{\hat{x},\mathrm{exp}}\sim \tilde{m}^{-1.86}$. (b) The integrated autocorrelation time\cite{autoremark} as a function of $\tilde{m}$. As for the exponential autocorrelation time, the  time required for two independent paths exhibits power-law behavior, $\tau_{\hat{x},\mathrm{int}}\sim \tilde{m}^{-1.85}$.}
\label{fig7}
\end{figure}

Figure~\ref{fig7} shows the dependence of the autocorrelation time of an observable $\hat{X}$ on the lattice spacing. For a given operator $\hat{O}$, we expect \cite{Gattringer,Janke} $\tau_O\sim\tilde{\xi}^z$ and $z\simeq 2$ for local updating algorithms.\cite{Montvay} Similar power-law behavior can be observed for the integrated autocorrelation time. We have chosen the number of lattice sites $N_{\tau}$ to be inversely related to the lattice spacing to keep the time $T$ in physical units constant. Hence, when moving toward the continuum, the  time increases as $\tau_O\tilde{\xi}^d$ in $d$ dimensions (in our case $d=1$).

Another way to express the computational difficulties in taking the continuum limit is that for fixed computational time, the exponential increase in the autocorrelation time as a function of the inverse lattice spacing causes an increase of the statistical errors, meaning that the reliability of results is limited for very small lattice spacings. A possible remedy is to change the updating strategy. The multigrid method, which employs intrinsically nonlocal updates, is an effective way to address large correlation lengths.

\section{\label{Overrelaxation}Over-relaxation}

\begin{figure}[t]
\includegraphics[width=7cm]{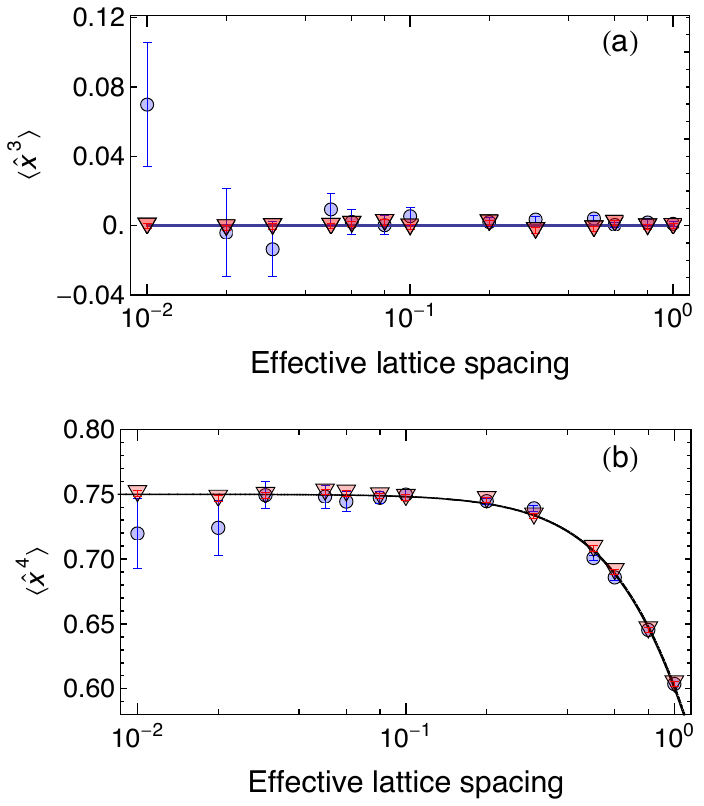}
\caption{Over-relaxed simulations  for the observables $\braket{\hat{x}^3}$ and $\braket{\hat{x}^4}$ are shown by the triangles ($N_{\tau}=120/\tilde{m}$; $N_{\mathrm{sep}}=300$). The reliability of the results is improved with no additional computational expense.}
\label{fig9}
\end{figure}

The goal of the over-relaxation method is to reduce autocorrelation times. To this end, a trial change $\tilde{x}_{i,\, \mathrm{trial} }$ far from the old value $\tilde{x}_i$, but involving small changes in the action, is proposed. Creutz \cite{Creutz2} and Brown and Woch \cite{Brown} suggest:
\begin{equation}
\label{xnewMC}
\tilde{x}_{i,\, \mathrm{trial} }=\frac{\tilde{x}_{i-1}+\tilde{x}_{i+1}}{1+{1\over2}\tilde{\omega}^2}-\tilde{x}_i\, .
\end{equation}
The variable
\begin{equation}\label{xnewSD}
\tilde{x}_{i,\, \mathrm{mid} }\equiv \frac{\tilde{x}_{i-1}+\tilde{x}_{i+1}}{2+\tilde{\omega}^2}
\end{equation}
minimizes the part of the action that depends on $\tilde{x}_i$. Thus, $\tilde{x}_{i,\, \mathrm{trial} }$ lies ``on the other side'' of this minimum for fixed $x_{i-1}$ and $x_{i+1}$. The update $\tilde{x}_i\to \tilde{x}_{i,\, \mathrm{trial} }$ is microcanonical, meaning that the action is constant under this change and no Metropolis accept/reject step is needed. The disadvantage of Eq.~(\ref{xnewMC}) is that the procedure is not applicable to actions for which $\tilde{x}_{i,\,\mathrm{mid}}$ cannot be found exactly, such as an anharmonic oscillator. Because the ratio of the kinetic to potential energy increases strongly in the continuum limit, we propose a trial change that preserves the kinetic part of the action:
\begin{equation}\label{xnewSD}
\tilde{x}_{i,\, \mathrm{kin} }=(\tilde{x}_{i-1}+\tilde{x}_{i+1})-\tilde{x}_i,
\end{equation}
followed by a standard accept/reject procedure.\cite{f5} Equation~(\ref{xnewSD}) can be used for any potential term. The only change in the pseudocode given in Ref.~\onlinecite{SI} is the value of $x_{\rm new}$. Usually all sweeps are Metropolis sweeps; for the over-relaxation routine used to create Fig.~\ref{fig9}, four in five Metropolis sweeps were exchanged for over-relaxed sweeps. The error bars are visibly smaller with over-relaxation. Because the CPU time needed for an over-relaxed sweep is comparable to that required for an ordinary Metropolis sweep, the accuracy of the measurements is significantly improved at constant computational cost.

\section{\label{sec7}Advanced topics}

\subsection{The ground state wave function}

\begin{figure}[t]
\includegraphics[width=7cm]{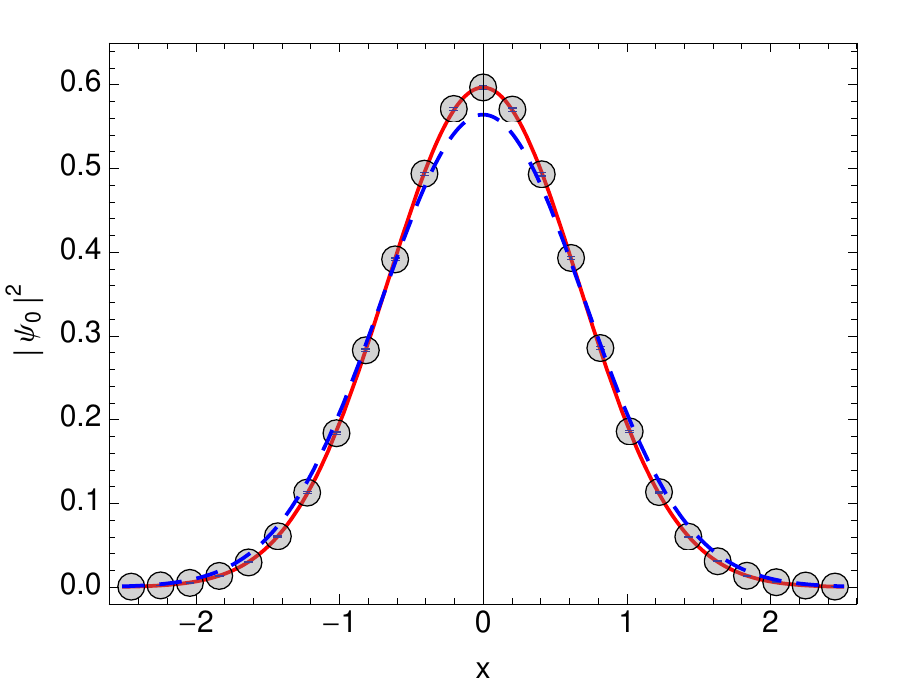}
\caption{(color online) Modulus squared of the ground state wave function for the harmonic oscillator. Measurements were based on $N=10^4$ paths with $\tilde{m}=\tilde{\omega}=1$, $N_{\tau}=120$ and bin size $\Delta x=0.1$. The error bars are smaller than the symbol size. The solid curve (red) represents the expression for $|\psi_0|^2$ on a discrete lattice for this choice of parameters, and the dashed curve (blue) is its continuum counterpart.}
\label{waveamp}
\end{figure}

Section~\ref{sec3} discussed how to determine any moment $\langle x^n\rangle$ of the ground state wave function of the harmonic oscillator. Here we point out that information about the complete ground state wave function (more precisely $|\psi_0(x)|^2$, because only the modulus squared of a wave amplitude can be measured in quantum mechanics) is directly accessible from the path integral. The squared modulus $|\psi_0(x)|^2$ is approximated by a histogram of the data for $x(\tau)$, which are available from the simulation. A histogram (over all configurations and all time slices within each configuration) is shown in Fig.~\ref{waveamp}. The data and full curve deviate from the dashed curve, which represents the continuum result (\ref{wavefunction}). Here the correction factor $(1+{1\over4}\tilde{\omega}^2)^{1/2}$ is apparent. This factor, seen in Eqs.~(\ref{x2}) and (\ref{x4}) and in Eq.~(\ref{xLadder}), accounts for the discreteness of Euclidean time. In the naive continuum limit $\tilde{\omega}\to0$, and the correction factor approaches one.

\subsection{Measuring energy differences}
The energy spectrum of the harmonic oscillator is $E_n=(n+\frac{1}{2})\omega$ for $n=0,1,2,\ldots$ ($\hbar=1$). It is possible to verify this energy spectrum by lattice simulations, up to the zero-point energy. In other words, we can measure the energy differences $\tilde{E}_m-\tilde{E}_n$ for any $m$ and $n$, but not the energies themselves.

To understand how, we revisit the two-point correlator $G(\Delta\tau)$ in Eq.~(\ref{normaltwopt}). It is instructive to rewrite it in terms of the ladder operators $\hat{a}$ and $\hat{a}^{\dagger}$ by substituting
\begin{equation}\label{xLadder}
\hat{x}=i \sqrt{\frac{1}{2\tilde{m}\tilde{\omega} (1+{1\over4}\tilde{\omega}^2)^{1/2}}}(\hat{a}-\hat{a}^{\dagger}),
\end{equation}
which follows from the definition of $\hat{a}$ and $\hat{a}^{\dagger}$ in terms of $\hat{x}$ and $\hat{p}$ (see Sec.~VII and Ref.~\onlinecite{SI} for details and the convention used). The operators $\hat{a}$ and $\hat{a}^{\dagger}$ act on the normalized eigenstates $|n\rangle$ of the number operator $\hat{a}^{\dagger}\hat{a}$, $
\hat{a}|0\rangle=0$ and $\left(\hat{a}^{\dagger}\right)^n |0\rangle =\sqrt{n!}|n\rangle
$.
With this substitution the two-point correlator takes the form
\begin{equation}
G(\Delta\tau)=
\frac{1}{2\tilde{m}\tilde{\omega}(1+{1\over4}\tilde{\omega}^2)^{1/2}}
\langle 0|\hat{a}(\tau)\hat{a}^{\dagger}(\tau+\Delta\tau)|0\rangle,
\end{equation}
because each one of the remaining three terms in the two-point correlator $G(\Delta\tau)$ with the substitution given in Eq.~(\ref{xLadder}) has an $\hat{a}^{\dagger}(\tau)$ operator acting on $\langle 0|$ to the left, and/or an $\hat{a}(\tau+\Delta\tau)$ acing on $|0\rangle $ to the right. After inserting a complete set of states $I=\sum |n\rangle \langle n|$ between the two ladder operators, we find that only the state $|1\rangle \langle 1|$ contributes. Finally, we write $\hat{A}(t)=\exp(\hat{H}t)\hat{A}\exp(-\hat{H}t)$ to find
\begin{align}\tilde{m}
G(\Delta\tau)&=
\frac{1}{2\tilde{m}\tilde{\omega}(1+{1\over4}\tilde{\omega}^2)^{1/2}}\langle 0|\hat{a}(\tau)|1\rangle  \langle 1|e^{-E_1\Delta\tau}\hat{a}^{\dagger}(\tau)e^{E_0\Delta\tau}|0\rangle \\
&=\frac{1}{2\tilde{m}\tilde{\omega}(1+{1\over4}\tilde{\omega}^2)^{1/2}}
e^{-(E_1-E_0)\Delta\tau}  \langle 0|\hat{a}(\tau)|1\rangle \langle 1|\hat{a}^{\dagger}(\tau)|0\rangle \\
&=e^{-(E_1-E_0)\Delta\tau}G(0),
\end{align}
which implies that the slope of $\log(G(\Delta\tau))$ \emph{versus} $\Delta\tau$ measures the energy difference $E_1-E_0$ (see Fig.~3).

To determine the difference $E_n-E_0$ for arbitrary $n$, we need to consider higher order multi-point functions. Equation~(\ref{xLadder}) implies that any function $\langle 0|\hat{x}(\tau_1)\ldots\hat{x}(\tau_{2n+1})|0\rangle$ with an odd number of $\hat{x}$ is zero. Accordingly, the first viable option is the 4-point function $\langle 0|\hat{x}(\tau_1)\hat{x}(\tau_2)\hat{x}(\tau_3)\hat{x}(\tau_4)|0\rangle$. For simplicity, we consider the restricted 4-point function with $\tau_1=\tau_2$ and $\tau_3=\tau_4$:
\begin{equation}
F(\Delta\tau)=
\langle 0|\hat{x}^2(\tau)\hat{x}^2(\tau+\Delta\tau)|0\rangle.
\label{f4}
\end{equation}
Squaring Eq.~(\ref{xLadder}) yields
\begin{equation}
\hat{x}^2=-\frac{1}{2\tilde{m}\tilde{\omega}(1+{1\over4}\tilde{\omega}^2)^{1/2}}(\hat{a}^2-\hat{a} \hat{a}^{\dagger}-\hat{a}^{\dagger} \hat{a}+\hat{a}^{\dagger2})
\end{equation}
and substituting this expression into Eq.~\eqref{f4} gives
\begin{align}
F(\Delta\tau)&=
\frac{1}{4\tilde{m}^2\tilde{\omega}^2(1+{1\over4}\tilde{\omega}^2)}\nonumber\\
&\times \langle 0|[\hat{a}^2-\hat{a} \hat{a}^{\dagger}](\tau)[-\hat{a} \hat{a}^{\dagger}+\hat{a}^{\dagger2}](\tau+\Delta\tau)|0\rangle,
\end{align}
because all other terms are zero. Only two of the four terms survive, and we find
\begin{align}
F(\Delta\tau)&=
\frac{1}{4\tilde{m}^2\tilde{\omega}^2(1+{1\over4}\tilde{\omega}^2)}\nonumber\\
&\times
[
\langle 0|\hat{a}^2(\tau)\hat{a}^{\dagger 2}(\tau+\Delta\tau)|0\rangle \nonumber\\
&+\langle 0|\hat{a} \hat{a}^{\dagger}(\tau) \hat{a} \hat{a}^{\dagger}(\tau+\Delta\tau)|0\rangle
].
\end{align}
As before, we insert a complete set of states $|n\rangle\langle n|$ between $\tau$ and $\tau+\Delta\tau$. It follows that the first term receives only a contribution from the $n=2$ state, and the second term only from the $n=0$ ground state. We again use the Heisenberg picture and find that the first term carries a time-dependence $\exp[-(E_2-E_0)\Delta \tau]$, and the second term carries no time-dependence. We thus define the connected correlator
\begin{align}
F_\mr{conn}(\Delta\tau)=
\langle 0|\hat{x}^2(\tau)\hat{x}^2(\tau+\Delta\tau)|0\rangle -
|\langle 0|\hat{x}^2(\tau)|0\rangle |^2,
\end{align}
which has the time-dependence
\begin{equation}
F_\mr{conn}(\Delta\tau)=
e^{-(E_2-E_0)\Delta\tau}F_\mr{conn}(0).
\end{equation}
We see that this correlator  decreases twice as fast (for the same parameters) as the 2-point function shown in Fig.~3. It is possible to generalize these considerations and to construct suitable combinations of $2n$-point functions that decrease as $\exp(-(E_n-E_0)\Delta\tau)$. Hence, we can determine the energy spectrum of the harmonic oscillator through  simulations. However, the higher multi-point functions will be noisier than the 2-point function.

\subsection{The anharmonic oscillator}
The anharmonic oscillator is interesting, not just as an application of the Monte Carlo Markov chain method developed here, but as a system where exact solutions are not available. The Rayleigh--Schr\"odinger perturbation is known\cite{bender69} to diverge, which has led to the development of approximate methods to estimate and place bounds on the energy levels of this system.\cite{montroll75}

The action of the quantum anharmonic oscillator is
\begin{equation}
{\tilde S}=\sum_{i=1}^{N_{\tau}} \frac{1}{2}\tilde{m}(\tilde{x}_{i+1}-\tilde{x}_i)^2 + \frac{1}{2}\tilde{m}\tilde{\omega}^2\tilde{x}_i^2+\frac{1}{4}\lambda\tilde{x}_i^4.
\end{equation}
All the techniques we have discussed can be applied to the anharmonic oscillator.
We have chosen the over-relaxation update $x_i \to \tilde{x}_i''$ in Eq.~(\ref{xnewSD}) such that it carries over to actions with an anharmonic term.  
That being said, the anharmonic oscillator is a much harder problem than the harmonic oscillator. We hope to come back to it in a future publication.

\section{Suggested problems}
As an exercise for the reader, we suggest developing the machinery in this paper step by step.

(a) Write code for a Metropolis sweep in a programming language of choice.
Study the pseudocode in Ref. \onlinecite{SI} to get started.

(b) Take $\tilde{m}=\tilde{\omega}=1$ and $N_{\tau}=120$, which is our coarsest lattice.
To thermalize the path, perform 100 Metropolis sweeps before saving it.
Carry out 12 Metropolis sweeps (written in a do/for loop) before saving the next path.
Repeat this procedure 10,000 times, always with $N_{\mathrm{sep}}=12$ (i.e., running the Metropolis sweep 12 times before saving the next path).
This makes for a grand total of 120,000 Metropolis sweeps in order to generate 10,000 paths.
Save each path externally.
As a sanity check, make a probability density diagram of all simulated positions (i.e., of $120\times10,000$ numbers) with bin size $\Delta x=0.1$.
A plot of these measurements should reproduce Fig. \ref{waveamp} (data).

(c) Measure a few observables based on the stored 10,000 configurations.
We suggest starting with $\langle\hat{x}\rangle$ and $\langle\hat{x}^2\rangle$; the former expectation value should be consistent with zero, the latter is given in Eqs. (\ref{x2}) and (\ref{aux}).
Modify your program so that it determines, after $N_{\mathrm{sep}}$ Metropolis sweeps, the average $\langle\hat{x}\rangle$ and the average $\langle\hat{x}^2\rangle$ of that path, and writes it to disk, rather than the path itself.
Extend your program so that it determines, from the same configurations, the intra-path averaged quantities $\langle\hat{x}^3\rangle$ and $\langle\hat{x}^4\rangle$, and writes these two numbers to disk, too.

(d) Repeat step (c) on progressively finer lattices, and combine the 10,000 numbers of $\langle\hat{x}^n\rangle$ into an ensemble average $\langle\hat{x}^n\rangle$ for every set of parameters.
The formulas for $\langle\hat{x}^2\rangle$ and $\langle\hat{x}^4\rangle$ as a function of the lattice spacing are given in Eqs. (\ref{x2})-(\ref{x4}).
Our values for $\tilde{m}=\tilde{\omega}$ and $N_{\tau}$ are stated in Table II.

(e) For each observable and choice of parameters, obtain an estimate of the asymptotic autocorrelation time $A_O$.
While $A_O$ can be determined through a fit, as in Eq. (63), a good way to get a feel for it is by using the matlab routine ``UWerr.m'' \cite{Wolff}, which is freely available online.
For a given dataset, it calculates an estimate of the asymptotic autocorrelation time and the associated statistical error.
Note that $N_{\mathrm{sep}}$ must increase for progressively finer lattices.
Is $N_{\mathrm{sep}}=N_{\tau}/10$ sufficient to keep the autocorrelation time to a reasonable magnitude
(i.e., small enough to have a sufficient number of independent ``bins'' on which to do a jackknife analysis)?

(f) Redo the calculations in (e) with a jackknife analysis (pseudocode is provided in Ref. \onlinecite{SI}).
The bin width should always be greater than the autocorrelation time of the dataset under consideration.

(g) Plot the results under (e) and (f) to visualize the reliability of the jackknife errors.

\section{Discussion}

How can we be confident in simulation data in the absence of predictions? The theory and data we have presented highlight two effects as the continuum limit is approached. The lattice-dependent theoretical answer converges to a limiting value; in the same limit, the noise-to-signal ratio blows up. Two main themes are of interest for the computation of path integrals on a lattice: the limiting answer and the trade-off between accuracy and required computer time. Because computer time requirements depend strongly on the autocorrelations present in the Markov chain, we should start with a systematic determination of the autocorrelation time $\tau_O$ as a function of the effective lattice spacing, keeping the product $N_{\tau}\Delta \tau$ fixed. For reliable estimates it is best to start with coarse lattices and high statistics and to determine the behavior of the autocorrelation function for progressively finer lattices. Once $\tau_O$ is established as a function of the effective lattice spacing, we can calculate the desired observable with an appropriate number of intermediate paths. The error bars should be determined with the jackknife procedure using an appropriate bin width.

The  answer in the continuum limit should reveal itself through a stable fit to the data obtained at the finest few lattice spacings, similar to our Fig. 8. If this stable fit is possible, we can have confidence in the continuum answer, even if there is no analytic answer available for comparison.

\begin{acknowledgments}
MJEW was supported through a Janet Watson scholarship from the Department of Earth Science and Engineering and a studentship in the Centre for Doctoral Training on Theory and Simulation of Materials funded by the EPSRC (EP/L015579/1), both at Imperial College London. SD acknowledges partial support by DFG through SFB-TR-55.
\end{acknowledgments}

\clearpage

\appendix 
\section{Elements of Dirac Bra-Ket Notation}

Dirac introduced bra-ket notation \cite{dirac39,dirac58} as a representation of states in a linear space that is free from the choice of coordinate, but enables the insertion of particular coordinates and the transformation between coordinate systems.  Dirac notation remains in widespread use today and current students will recognize the connection to the linear algebra of vectors and matrices. In this section we review the main principles of bras and kets in quantum  mechanics.

\begin{table}[b]
\begin{ruledtabular}
\begin{tabular}{ l l l l}
orthonormal& $|\phi_1\rangle, |\phi_1\rangle,\ldots$ &orthonormal& $\bm{e}_1,\bm{e}_2,\ldots,\bm{e}_n$\\
basis & & basis &\\
\noalign{\vskip12pt}
ket & $|\varphi_a\rangle=\sum_ia_i|\phi_i\rangle$ & column vector & $\bm{a}=\left(\begin{array}{c}a_1\\ \noalign{\vskip3pt}a_2\\ \vdots\\ a_n\end{array}\right)=\sum_ia_i\bm{e}_i$\\
\noalign{\vskip12pt}
bra & $\langle\varphi_a|=\sum_ia_i^\ast\langle\phi_i|$ & row vector & $\bm{a}^\dagger=(a_1^\ast\hskip3pt a_2^\ast\cdots a_n^\ast)=\sum_ia_i^\ast\bm{e}_i^\dagger$\\
\noalign{\vskip12pt}
inner product &  $\langle\varphi_a|\varphi_b\rangle=\sum_{i,j} a_i^\ast b_j\langle\phi_i|\phi_j\rangle$ 
& scalar product & $\bm{a}^\dagger\bm{b}=(a_1^\ast\hskip3pt a_2^\ast\cdots a_n^\ast)\left(\begin{array}{c}b_1\\ \noalign{\vskip3pt}b_2\\ \vdots\\ b_n\end{array}\right)=\sum_ia_i^\ast b_i$\\
& $=\sum_ia_i^\ast b_i$ & \\
\noalign{\vskip12pt}
operator & $\hat{A}=\sum_{i,j}A_{ij}|\phi_i\rangle\langle\phi_j|$ & matrix & {\sf A}=$\left(\begin{array}{cccc}A_{11}&\hskip3pt A_{12}&\cdots& A_{1n}\\ \noalign{\vskip3pt}
A_{21}&\hskip3pt A_{22}&\cdots& A_{2n}\\ \vdots & \vdots &\ddots &\cdots\\
\noalign{\vskip3pt}
A_{n1} & A_{n2} &\cdots & A_{nn}\end{array}\right)$\\
\noalign{\vskip12pt}
dyadic product & $|\varphi_a\rangle\langle\varphi_b|=\sum_{i,j}a_ib_j^\ast|\phi_i\rangle\langle\phi_j|$ & dyadic product & $\bm{a}\bm{b}^\dagger=\left(\begin{array}{c}a_1\\ \noalign{\vskip3pt}a_2\\ \vdots\\ a_n\end{array}\right)(b_1^\ast\hskip3pt b_2^\ast\cdots b_n^\ast)$\\
\noalign{\vskip12pt}
& & & \hskip0.7cm$=\left(\begin{array}{cccc}a_1b_1^\ast&\hskip3pt a_1b_2^\ast&\cdots& a_1b_n^\ast\\ \noalign{\vskip3pt}
a_2b_1^\ast&a_2b_2^\ast&\cdots& a_2b_n^\ast\\ \vdots & \vdots &\ddots &\cdots\\
\noalign{\vskip3pt}
a_nb_1^\ast & a_nb_2^\ast &\cdots & a_nb_n^\ast\end{array}\right)$
\end{tabular}
\end{ruledtabular}
\caption{\label{tableS1}Correspondence between various quantities in Dirac bra-ket notation and their corresponding matrix constructions in an $n$-dimensional complex vector space.}
\end{table}

%%%%%%%%%%%%%%%%%%%%%%%%%%%%%%%%%%%%%%%%%%%%%%%%%

\subsection{Bras and Kets}

The fundamental entities of Dirac's bra-ket notation are bras and kets.  The ket, denoted as $|\,\cdot\,\rangle$,  signifies how a quantum state is characterized.  For example, $|\bm{p}\rangle$ signifies a state with momentum $\bm{p}$, $|\bm{x}\rangle$ a state at coordinate $\bm{x}$, and $|\psi\rangle$ a system in state $\psi$.  If the context is apparent, the entries of a ket can be quantum numbers.  For example, $ |n\rangle$ signifies a system in the $n$th quantum state, and  for the hydrogen atom, $|n\ell m\rangle$ signifies a state with principal quantum number $n$, angular momentum quantum number $\ell$, and magnetic quantum number $m$.  The ket can also represent the initial state of a system before a transition.

The bra, signified by $\langle\,\cdot\,|$,  contains the representation of a ket.  For example, $\langle \bm{x}|\psi\rangle$ is the amplitude that a system in state $\psi$ is located at $\bm{x}$:~$\langle\bm{x}|\psi\rangle=\psi(\bm{x})$. Similarly, $\langle\bm{p}|\psi\rangle$ is the amplitude of a system in state $\psi$ with momentum $\bm{p}$.  The quantity $\langle x|n\rangle=\psi_n(x)$ is the coordinate representation of the $n$th eigenstate of a one-dimensional system, such as a particle in an infinite square well or the harmonic oscillator.  The bra can also represent the final state of a system after a transition.

\subsection{Bra-Ket and Ket-Bra Pairs}

A bra-ket pair $\langle\,\cdot\,|\,\cdot\,\rangle$ is analogous to a projection, in which the entry of the ket is projected onto the entry of the bra.  For example, $\langle\phi|\psi\rangle$ is the projection of $\psi$ onto $\phi$; that is, the amount of the state $\psi$ contained in the state $\phi$, yielding what is known as an overlap integral when evaluated in coordinate space (see below).  More generally, $\langle x|\psi\rangle=\psi(x)$ is the projection of $\psi$ onto the coordinate $x$ and $\langle p|\psi\rangle$ is the projection of $\psi$ onto the momentum $p$.

Because state vectors are, in general, complex quantities, projections are also complex.  A special case is the projection of a state vector onto itself:~$\langle\psi|\psi\rangle$, which is the overlap of the state $\psi$ with itself and is the analogue of the magnitude of a complex vector, which must be a real number.  This analogy can be guaranteed for general state vectors only if $\langle\,\cdot\,|=|\,\cdot\,\rangle^\dagger$, where $|\,\cdot\,\rangle^\dagger$ is the Hermitian conjugate of $|\,\cdot\,\rangle$.

The ket-bra product $|\,\cdot\,\rangle\langle\,\cdot\,|$ is a projection operator.  For example, operating with $|\phi\rangle\langle\phi|$ on $|\psi\rangle$ yields $|\phi\rangle\langle\phi|\psi\rangle$, which is the projection of $\psi$ onto $\phi$ multiplied by $|\phi\rangle$: the state vector $|\psi\rangle$ projected onto $|\phi\rangle$.  The projection operator can be extended to any number of components.  Consider, for example, three-dimensional Euclidean space, whose unit basis vectors along the $x$-, $y$, and $z$-axes are $\bm{i}$, $\bm{j}$, and $\bm{k}$, respectively.  The projection operator for this space is
\begin{equation}\label{ProjectionOperator}
|\bm{i}\rangle\langle\bm{i}|+|\bm{j}\rangle\langle\bm{j}|+|\bm{k}\rangle\langle\bm{k}|\, .
\end{equation}
Operating on any three-dimensional vector $|\bm{u}\rangle$ produces
\begin{equation}
|\bm{i}\rangle\langle\bm{i}|\bm{u}\rangle+|\bm{j}\rangle\langle\bm{j}|\bm{u}\rangle+|\bm{k}\rangle\langle\bm{k}|\bm{u}\rangle\, .
\end{equation}
This  application of the projection operator (\ref{ProjectionOperator}) yields the  representation of any three-dimensional vector in terms of its Cartesian coordinates. Therefore, the sum is  equal to $|\bm{u}\rangle$:
\begin{equation}
|\bm{i}\rangle\langle\bm{i}|\bm{u}\rangle+|\bm{j}\rangle\langle\bm{j}|\bm{u}\rangle+|\bm{k}\rangle\langle\bm{k}|\bm{u}\rangle=|\bm{u}\rangle\, .
\label{eq3}
\end{equation}
 Because  we have chosen our test vector arbitrarily,  we can write
\begin{equation}
|\bm{i}\rangle\langle\bm{i}|+|\bm{j}\rangle\langle\bm{j}|+|\bm{k}\rangle\langle\bm{k}|=\mathbb{1}\, ,
\label{eq3a}
\end{equation}
where $\mathbb{1}$ is a $3\times3$ unit matrix.  Operating both sides of this equation on any vector $|\bm{u}\rangle$ yields Eq.~(\ref{eq3}).

Equation~(\ref{eq3a}) is known as a {\it completeness relation} because every vector can be represented as the sum of three Cartesian components.  As a counterexample,  $|\bm{i}\rangle\langle\bm{i}|+|\bm{j}\rangle\langle\bm{j}|$ is not a completeness relation for all three-dimensional vectors because the $z$-component is not included in this sum. It is, however, complete for vectors in the $x$-$y$ plane.

The same principles apply to an infinite set of functions, typically eigenfunctions of a Hamiltonian:
\begin{align}
\label{eq4}
\sum_{n=1}^\infty |n\rangle\langle n|&=\mathbb{1}\, ,\\
\sum_{n=1}^\infty\sum_{\ell=1}^{n-1}\sum_{n=-\ell}^\ell |n\ell m\rangle\langle n\ell m|&=\mathbb{1}\, ,
\label{eq5}
\end{align}
where, in this case, $\mathbb{1}$ is an infinite-dimensional unit matrix.  In the case of continuous variables, the summations become integrals:
\begin{align}
\label{eq6}
\int |x\rangle\langle x|\,dx&=\hat{\mathbb{1}}\, ,\\
\int |p\rangle\langle p|\,dp&=\hat{\mathbb{1}}\, ,
\label{eq7}
\end{align}
in which $\hat{\mathbb{1}}$ is the unit operator.  The meaning of Eqs.~(\ref{eq4})--(\ref{eq7}) is analogous to the vector case (\ref{eq3}). In each case a function can be expressed uniquely as a linear combination of an appropriate basis. For example, by applying Eq.~(\ref{eq6}) to a state ket $|\psi\rangle$,
\begin{equation}
\int_{x\in D} |x\rangle\langle x|\psi\rangle\,dx=|\psi\rangle\, ,
\label{eq8}
\end{equation}
which expresses the fact that the coordinate of the system in state $\psi$ is somewhere in the domain $D$ of allowed coordinates, which could be the real line for a free particle in one dimension, or a finite interval if the particle is confined to a square well.  Then, operating with the bra $\langle x^\prime|$, we obtain
\begin{equation}
\int_{x\in D}\langle x^\prime |x\rangle\langle x|\psi\rangle\,dx=\langle x^\prime|\psi\rangle=\psi(x^\prime)\, . \label{rhs}
\end{equation}
The right-hand side of Eq.~\eqref{rhs} is the projection of $|\psi\rangle$ onto $\langle x^\prime|$ or, in conventional quantum mechanical language, the amplitude of $\psi$ at $x^\prime$.   That the left-hand side yields the same quantity can be seen by using the fact that $\langle x^\prime |x\rangle=\delta(x-x^\prime)$, where $\delta$ is the Dirac delta function, which is defined by
\begin{equation}
\int\delta(x)\,dx=1\, ,
\end{equation}
such that
\begin{equation}
\int f(x^\prime)\delta(x-x^\prime)\,dx=f(x)\, .
\end{equation}
For our purposes, the  relation $\langle x^\prime |x\rangle=\delta(x-x^\prime)$ means that there is no overlap in these coordinate states.   

The completeness relation (\ref{eq6}) can also be used to provide a coordinate representation of $\langle\psi|\psi\rangle$:
\begin{equation}
\langle\psi|\psi\rangle=\!\int_{x\in D}\langle\psi|x\rangle\langle x|\psi\rangle\,dx=\!\int_{x\in D}\psi^\ast(x)\psi_x(x)\,dx\, ,
\end{equation}
which is the normalization integral for $\psi$.  Notice that in the second equality, Eq.~(\ref{eq6}) has been inserted between the bra and ket.

%%%%%%%%%%%%%%%%%%%%%%%%%%%%%%%%%%%%%%%%%%%%%%%%%

\subsection{Momentum Operator in Coordinate Space}

The canonical commutation relation between the coordinate and momentum operators is
\vspace{-12pt}
\begin{equation}
[\hat{x},\hat{p}]=\hat{x}\,\hat{p}-\hat{p}\,\hat{x}=i\hbar\, .
\end{equation} 
Taking matrix elements between states $\phi$ and $\psi$ yields
\begin{equation}
\langle\phi|[\hat{x},\hat{p}]|\psi\rangle=i\hbar\langle\phi|\psi\rangle\, .
\label{eq14}
\end{equation}
The left-hand side can be written by using Eq.~(\ref{eq6}) twice:
\begin{subequations}
\begin{align}
\int dx\!\int dx^\prime\langle\phi| x\rangle&\langle x|[\hat{x},\hat{p}]| x^\prime\rangle\langle x^\prime|\psi\rangle =
\!\int dx\!\int dx^\prime\,\langle\phi| x\rangle\langle x|\hat{x}\hat{p}-\hat{p}\hat{x}| x^\prime\rangle\langle x^\prime|\psi\rangle \\
&=\!\int dx\!\int dx^\prime \phi^\ast(x)\big(x\langle x|\hat{p}| x^\prime\rangle-\langle x|\hat{p}|x^\prime\rangle x^\prime\big)
\psi(x')\\
&=\!\int \phi^\ast(x)\psi(x)\, dx\, ,
\end{align}
\end{subequations}
where we have used the fact that $\hat{x}$ is a Hermitian operator, so the Hermitian conjugate of $\hat{x}|x\rangle=x|x\rangle$ is $\langle x|\hat{x}=\langle x|x$.  We have also used Eq.~(\ref{eq6}) to obtain the coordinate representation of the right-hand side of Eq.~(\ref{eq14}).  The quantity to be determined is $\langle x|\hat{p}| x^\prime\rangle$.  The last equality necessitates eliminating one of the integrals in the penultimate line, which  is accomplished by the delta function, $\delta(x-x^\prime)$.  Therefore, the equation for $\langle x|\hat{p}| x^\prime\rangle$ reduces to
\begin{equation}
x\langle x|\hat{p}| x\rangle\psi(x)-\langle x|\hat{p}|x\rangle\big[x \psi(x)\big]=i\hbar \psi(x)\, .
\end{equation}
Hence, we see that $\langle x|\hat{p}| x^\prime\rangle$ must be a differential operator:
\begin{equation}
\langle x|\hat{p}| x^\prime\rangle=-i\hbar\,\delta(x-x^\prime)\,{d\over dx}\, .
\end{equation}

With this result, we can determine the transformation between coordinate and momentum bases, $\langle x|p\rangle$.  We begin with $\langle x|\hat{p}|p\rangle$, which we evaluate in two ways:
\begin{align}
\langle x|\hat{p}|p\rangle&=p\langle x|p\rangle\, ,\\
\langle x|\hat{p}|p\rangle&=\!\int\langle x|\hat{p}|x^\prime\rangle\langle x^\prime|p\rangle\,dx=-i\hbar{d\langle x|p\rangle\over dx}\, ,
\end{align}
Equating the two right-hand sides yields the differential equation,
\begin{equation}
-i\hbar{d\langle x|p\rangle\over dx}=p\langle x|p\rangle\, ,
\end{equation}
whose solution is
\begin{equation}
\langle x|p\rangle={1\over\sqrt{2\pi\hbar}}\exp\bigg({ipx\over\hbar}\bigg)\, ,
\label{eq22}
\end{equation}
where the prefactor is chosen to ensure that $\langle x|x^\prime\rangle=\delta(x-x^\prime)$:
\begin{subequations}
\begin{align}
\langle x|x^\prime\rangle&=\!\int_{-\infty}^\infty\langle x|p\rangle\langle p|x^\prime\rangle\,dp=
{1\over2\pi \hbar}\!\int_{-\infty}^\infty\exp\bigg({ipx\over\hbar}\bigg)\exp\bigg(-{ipx^\prime\over\hbar}\bigg)\,dp \\
&={1\over2\pi \hbar}\!\int_{-\infty}^\infty\exp\bigg[{ip\over\hbar}(x-x^\prime)\bigg]\,dp=
{1\over2\pi}\!\int_{-\infty}^\infty e^{ik(x-x^\prime)}\,dp=\delta(x-x^\prime)\, .
\end{align}
\end{subequations}
We have used the relation $p=\hbar k$ to transform the integration variable from $p$ to $k$.

%%%%%%%%%%%%%%%%%%%%%%%%%%%%%%%%%%%%%%%%%%%%%%%%%

\section{Derivation of the Propagator}

The propagator is
\begin{equation}
\langle x_i,t_i+t|x_i,t_i\rangle=\!\int_{-\infty}^\infty\langle x_f|p\rangle\langle p|e^{-i\mathcal{\hat{H}}t/\hbar}|x_i\rangle\,dp\, .
\label{eq24}
\end{equation}
In the limit of small $t\equiv \delta t$, we can expand the exponential on the right-hand side and retain terms only to first order in $\delta t$:
\begin{subequations}
\begin{align}
\langle p|e^{-i\mathcal{\hat{H}}\delta t/\hbar}|x_i\rangle&=\bigg\langle p\,\bigg|\mathbb{1}-{i\mathcal{\hat{H}}\delta t\over\hbar}+{\cal O}(\delta t^2)\bigg|\,x_i\bigg\rangle \\
&=\langle p|\mathbb{1}|x_i\rangle-{i\delta t\over\hbar}\langle p|\mathcal{\hat{H}}|x_i\rangle+{\cal O}(\delta t^2)\, .
\label{eq25}
\end{align}
\end{subequations}
The first integral on the right-hand side can be written as 
\begin{equation}
\langle p|\mathbb{1}|x_i\rangle=\mathbb{1}\langle p|x_i\rangle\, .
\label{eq26}
\end{equation}
To evaluate the matrix element in the second term on the right-hand side, we first write
\begin{equation}
\langle p|\mathcal{\hat{H}}|x_i\rangle=\bigg\langle p\,\bigg|{\hat{p}^2\over2m}+V(\hat{x})\bigg|\, x_i\bigg\rangle=
{1\over2m}\langle p|\hat{p}^2|x_i\rangle + \langle p|V(\hat{x})|x_i\rangle\, .
\label{eq27}
\end{equation}
The momentum operator is Hermitian and so can operate from the left or right.  The matrix element in the first term on the right-hand side can thereby be written as
\begin{equation}
\langle p|\hat{p}^2|x_i\rangle=\langle p|p^2|x_i\rangle=p^2\langle p|x_i\rangle\, .
\end{equation}
The second term on the right-hand side of Eq.~(\ref{eq27}) can be written as
\begin{equation}
\langle p|V(\hat{x})|x_i\rangle=\langle p|V(x_i)|x_i\rangle=V(x_i)\langle p|x_i\rangle\, .
\label{eq29}
\end{equation}
Substituting Eqs.~(\ref{eq26})--(\ref{eq29}) into Eq.~(\ref{eq25}) gives, 
\begin{subequations}
\begin{align}
\langle p|e^{-i\mathcal{\hat{H}}\delta t/\hbar}|x_i\rangle&=\bigg\{\mathbb{1}-{i\delta t\over\hbar}\bigg[{p^2\over2m}+V(x_i)\bigg]\bigg\}\langle p|x_i\rangle+{\cal O}(\delta t^2)\\
&=\exp\bigg\{-{i\delta t\over\hbar}\bigg[{p^2\over2m}+V(x_i)\bigg]\bigg\}\langle p|x_i\rangle+{\cal O}(\delta t^2)\, ,
\end{align}
\end{subequations}
whereupon the right-hand side of Eq.~(\ref{eq24}) becomes
\begin{equation}
\int_{-\infty}^\infty\langle x_f|p\rangle\langle p|e^{-i\mathcal{\hat{H}}t/\hbar}|x_i\rangle\,dp=
\!\int_{-\infty}^\infty\langle x_f|p\rangle\exp\bigg\{-{i\delta t\over\hbar}\bigg[{p^2\over2m}+V(x_i)\bigg]\bigg\}\langle p|x_i\rangle\, dp\, ,
\label{eq31}
\end{equation}
where we have dropped the reference to ${\cal O}(\delta t^2)$ corrections.  From Eq.~(\ref{eq22}), we have 
\begin{equation}
\langle x_f|p\rangle={1\over\sqrt{2\pi\hbar}}\exp\bigg({ipx_f\over\hbar}\bigg)\, ,\qquad
\langle p |x_i\rangle={1\over\sqrt{2\pi\hbar}}\exp\bigg(-{ipx_i\over\hbar}\bigg)\, ,
\end{equation}
which, when substituted into Eq.~(\ref{eq31}) yields
\begin{subequations}
\begin{align}
&\!\int_{-\infty}^\infty\langle x_f|p\rangle\exp\bigg\{-{i\delta t\over\hbar}\bigg[{p^2\over2m}+V(x_i)\bigg]\bigg\}\langle p|x_i\rangle\, {dp\over2\pi\hbar}\\
&=\exp\bigg[-{i\delta t\over\hbar}V(x_i)\bigg]\!\int_{-\infty}^\infty\exp\bigg({ipx_f\over\hbar}\bigg)\exp\bigg(-{ip^2\delta t\over2m\hbar}\bigg)\exp\bigg(-{ipx_i\over\hbar}\bigg)\,{dp\over2\pi\hbar}\\
&=\exp\bigg[-{i\delta t\over\hbar}V(x_i)\bigg]\!\int_{-\infty}^\infty\exp\bigg\{-{i\delta t\over2m\hbar}\bigg[p^2-2mp\bigg({x_f-x_i\over\delta t}\bigg)\bigg]\bigg\}\,{dp\over2\pi\hbar}\, .
\label{eq33}
\end{align}
\end{subequations}
The quantity in square brackets in the argument of the exponential can be written as
\begin{equation}
p^2-2mp\bigg({x_f-x_i\over\delta t}\bigg)=\bigg[p-m\bigg({x_f-x_i\over\delta t}\bigg)\bigg]^2-m^2\bigg({x_f-x_i\over\delta t}\bigg)\, ,
\end{equation}
which enables the right-hand side of Eq.~(\ref{eq33}) to be written as
\begin{align}
&\exp\bigg[-{i\delta t\over\hbar}V(x_i)\bigg]\exp\bigg\{{i\delta t\over\hbar}\bigg[{m\over2}\bigg({x_f-x_i\over\delta t}\bigg)^2\bigg]\bigg\}\nonumber\\
& \quad \times\! \int_{-\infty}^\infty\exp\bigg\{-{i\delta t\over2m\hbar}\bigg[p-m\bigg({x_f-x_i\over\delta t}\bigg)\bigg]^2\bigg\}\,{dp\over2\pi\hbar}\, .
\label{eq35}
\end{align}
To evaluate the integral, we first transform the momentum variable $p$ to a new momentum variable $\tilde{p}$ according to
\begin{equation}
\tilde{p}=p-m\bigg({x_f-x_i\over\delta t}\bigg)\, ,
\end{equation}
which is a simple (real) translation.  The integral becomes
\begin{equation}
\int_{-\infty}^\infty\exp\bigg(-{i \,\delta t\, \tilde{p}^2\over2m\hbar}\bigg)\,{d\tilde{p}\over2\pi\hbar}\, .
\end{equation}
We next rescale $\tilde{p}$,
\begin{equation}
s=\bigg({\delta t\over2m\hbar}\bigg)^{\! 1\over2}\tilde{p}\, ,
\end{equation}
to obtain
\begin{equation}
\bigg({2m\hbar\over\delta t}\bigg)^{1\over 2}\!\int_{-\infty}^\infty e^{-is^2}\,ds=\bigg({2m\hbar\over\delta t}\bigg)^{1\over 2}\bigg({\pi\over i}\bigg)^{1\over2}=\bigg({2\pi m\hbar\over i \delta t}\bigg)^{1\over 2}\, .
\label{eq39}
\end{equation}
Hence, by combining Eqs.~(\ref{eq31}),  (\ref{eq33}), (\ref{eq35}), and (\ref{eq39}), we obtain the propagator at short times as
\begin{subequations}
\begin{align}
\langle x_f,t_i+\delta t|x_i,t_i\rangle&={1\over2\pi\hbar}\bigg({2\pi m\hbar\over i \delta t}\bigg)^{1\over 2}
\exp\bigg\{{i\delta t\over\hbar}\bigg[{m\over2}\bigg({x_f-x_i\over\delta t}\bigg)^2\bigg]\bigg\}\exp\bigg[-{i\delta t\over\hbar}V(x_i)\bigg] \\
&=\bigg({m\over2\pi i\hbar  \delta t}\bigg)^{1\over 2}
\exp\bigg\{{i\delta t\over\hbar}\bigg[{m\over2}\bigg({x_f-x_i\over\delta t}\bigg)^2-V(x_i)\bigg]\bigg\}\, .
\end{align}
\end{subequations}
By interpreting 
\begin{equation}
{x_f-x_i\over\delta t}\equiv{\delta x\over\delta t}
\end{equation}
as a discrete velocity, the argument of the exponential is a discrete Lagrangian, ${\cal L}=T-V$, where $T$ and $V$ are the discrete kinetic and potential energies, respectively:
\begin{equation}
{\cal L}={m\over2}\bigg({x_f-x_i\over\delta t}\bigg)^2-V(x_i)\, .
\end{equation}
The short-time propagator can then be written in a particularly compact form as
\begin{equation}
\langle x_f,t_i+\delta t|x_i,t_i\rangle=\bigg({m\over2\pi i\hbar  \delta t}\bigg)^{1\over 2}\exp\bigg({i{\cal L}\delta t\over\hbar}\bigg)\, .
\end{equation}

%%%%%%%%%%%%%%%%%%%%%%%%%%%%%%%%%%%%%%%%%%%%%%%%%

\section{Derivation of the Propagator in Imaginary Time}

The calculation of the imaginary-time propagator proceeds by essentially the same steps as the real-time propagator in the preceding section.  The propagator for short imaginary times is
\begin{equation}
\langle x_f|e^{-\mathcal{\hat{H}}\delta\tau/\hbar}|x_i\rangle=\!\int_{-\infty}^\infty\langle x_f|p\rangle \langle p|e^{-\mathcal{\hat{H}}\delta\tau/\hbar}|x_i\rangle\,dp\, .
\label{eq44}
\end{equation}
Expanding the exponential and retaining terms only to first order in $\delta\tau$ yields
\begin{subequations}
\begin{align}
\langle p|e^{-\mathcal{\hat{H}}\delta\tau/\hbar}|x_i\rangle
&=\bigg\langle p\,\bigg|\mathbb{1}-{\mathcal{\hat{H}}\delta\tau\over\hbar}+\cal{O}(\delta\tau^2)\bigg|\,x_i\bigg\rangle=\langle\mathbb{1}|x_i\rangle-{\delta\tau\over\hbar}\langle p|\mathcal{\hat{H}}|x_i\rangle+{\cal O}(\delta\tau^2) \\
&=\mathbb{1}\langle p|x_i\rangle-{\delta\tau\over\hbar}\bigg\langle p\,\bigg|{\hat{p}^2\over2m}+V(\hat{x})\bigg|\,x_i\bigg\rangle+{\cal O}(\delta\tau^2)\\
&=\mathbb{1}\langle p|x_i\rangle-{\delta\tau\over\hbar}\bigg[{p^2\over2m}+V(x_i)\bigg]\langle p|x_i\rangle+{\cal O}(\delta\tau^2) \\
&=\bigg\{\mathbb{1}-{\delta\tau\over\hbar}\bigg[{p^2\over2m}+V(x_i)\bigg]\bigg\}\langle p|x_i\rangle+{\cal O}(\delta\tau^2)\\
&=\exp\bigg\{-{\delta\tau\over\hbar}\bigg[{p^2\over2m}+V(x_i)\bigg]\bigg\}\langle p|x_i\rangle+{\cal O}(\delta\tau^2)\, .
\end{align}
\end{subequations}
By substituting this result into Eq.~(\ref{eq44}), invoking Eq.~(\ref{eq22}), and removing the explicit reference to the ${\cal O}(\delta\tau^2)$ corrections, we obtain
\begin{subequations}
\begin{align}
\int_{-\infty}^\infty&\langle x_f|p\rangle \langle p|e^{-\mathcal{\hat{H}}\delta\tau/\hbar}|x_i\rangle\,dp=\!\int_{-\infty}^\infty\langle x_f|p\rangle\exp\bigg\{-{\delta\tau\over\hbar}\bigg[{p^2\over2m}+V(x_i)\bigg]\bigg\}\langle p|x_i\rangle\,dp \\
&=\exp\bigg[-{\delta\tau\over\hbar}V(x_i)\bigg]\!\int_{-\infty}^\infty\exp\bigg({ipx_f\over\hbar}\bigg)\exp\bigg(-{p^2\delta\tau\over2m\hbar}\bigg)\exp\bigg(-{ipx_i\over\hbar}\bigg)\,{dp\over2\pi\hbar} \\
&=\exp\bigg[-{\delta\tau\over\hbar}V(x_i)\bigg]\!\int_{-\infty}^\infty\exp\bigg[{ip(x_f-x_i)\over\hbar}-{p^2\delta\tau\over2m\hbar}\bigg]\,{dp\over2\pi\hbar}\\
&=\exp\bigg[-{\delta\tau\over\hbar}V(x_i)\bigg]\!\int_{-\infty}^\infty\exp\bigg\{-{\delta\tau\over2m\hbar}\bigg[p^2-2mip\bigg({x_f-x_i\over\delta\tau}\bigg)\bigg]\bigg\}\,{dp\over2\pi\hbar}\, .
\label{eq46}
\end{align}
\end{subequations}
We again proceed to evaluate this integral by first completing the square of the terms contained within the square brackets:
\begin{equation}
p^2-2mip\bigg({x_f-x_i\over\delta\tau}\bigg)=\bigg[p-im\bigg({x_f-x_i\over\delta\tau}\bigg)\bigg]^2+m^2\bigg({x_f-x_i\over\delta\tau}\bigg)^2\, .
\end{equation}
The right-hand side of Eq.~(\ref{eq46}) thereby becomes
\begin{equation}
\exp\bigg\{-{\delta\tau\over\hbar}\bigg[{m\over2}\bigg({x_f-x_i\over\delta\tau}\bigg)^2+V(x_i)\bigg]\bigg\}\!\int_{-\infty}^\infty\exp\bigg\{-{\delta\tau\over2m\hbar}\bigg[p-im\bigg({x_f-x_i\over\delta}\bigg)\bigg]^2\bigg\}\,{dp\over2\pi\hbar}\, .
\end{equation}

To evaluate this integral, we first transform to a shifted momentum $\tilde{p}$ such that
\begin{equation}
\tilde{p}=p-im\bigg({x_f-x_i\over\delta\tau}\bigg)\equiv p-i\Delta\, ,
\end{equation}
so $p=\tilde{p}+i\Delta$ and the integral becomes
\begin{equation}
\int_{-\infty-i\Delta}^{\infty-i\Delta}\exp\bigg(-{\tilde{p}^2\delta \tau\over2m\hbar}\bigg)\,{dp\over2\pi\hbar}\, .
\end{equation}
Finally, we  transform to a variable $s$:
\begin{equation}
s^2={\tilde{p}^2\delta\tau\over2m\hbar}\quad\longrightarrow\quad \tilde{p}=s\bigg({2m\hbar\over\delta\tau}\bigg)^{1\over2}\, ,
\end{equation}
and the integral becomes
\begin{equation}
\bigg({m\over2\pi^2\hbar\delta\tau}\bigg)^{1\over2}\!\int_{-\infty-i\Delta^\prime}^{\infty-i\Delta^\prime}e^{-s^2}\, ds\, ,
\label{eq52}
\end{equation}
where
\begin{equation}
\Delta^\prime=\bigg({m\delta\tau\over2\hbar}\bigg)^{1\over2}\bigg({x_f-x)i\over\delta\tau}\bigg)\, .
\end{equation}
Thus we obtain
\begin{equation}
\langle x_f|e^{-\mathcal{\hat{H}}\delta\tau/\hbar}|x_i\rangle=\bigg({m\over2\pi\hbar\delta\tau}\bigg)^{1\over2}
\exp\bigg\{-{\delta\tau\over\hbar}\bigg[{m\over2}\bigg({x_f-x_i\over\delta\tau}\bigg)^2+V(x_i)\bigg]\bigg\}\, .
\end{equation}
If we define the ``Lagrangian'' $\tilde{\cal L}$ as
\begin{equation}
\tilde{\cal L}={m\over2}\bigg({x_f-x_i\over\delta\tau}\bigg)^2+V(x_i)\, ,
\end{equation}
then the propagator at short imaginary times can be written as
\begin{equation}
\langle x_f|e^{-\mathcal{\hat{H}}\delta\tau/\hbar}|x_i\rangle=\bigg({m\over2\pi\hbar\delta\tau}\bigg)^{1\over2}
\exp\bigg(-{\tilde{\cal L}\delta\tau\over\hbar}\bigg)\, .
\end{equation}

\section{Derivation of the relation $\left<\hat{x}^4\right>=3\left<\hat{x}^2\right>^2$}
\label{App:Creutz}
In this section we prove  the relation $\left<\hat{x}^4\right>=3\left<\hat{x}^2\right>$ for the quantum harmonic oscillator. The derivation of $\left<\hat{x}^2\right>$ is based on Appendix C in  Creutz \& Freedman,\cite{Creutz}  and the formulation of $\left<\hat{x}^4\right>$ is an extension of  that  work. The results are valid for any lattice spacing and can be used to verify lattice-based calculations. The parameter $N$ in this section corresponds to our number of lattice indices $N_{\tau}$; we eliminated the lattice spacing $a$ by  making the variables dimensionless.

\subsection{The relation $\left<\hat{x}^4\right>=3\left<\hat{x}^2\right>^2$ in a free field theory}\label{sec1}
This section shows that the relation $\left<\hat{x}^4\right>=3\left<\hat{x}^2\right>$ holds true in any free field theory. In particular, we can choose the trajectory $x(t)$ of the quantum harmonic oscillator as our ``field''. 
The generating functional for a Gaussian free field $\phi(x)$ is given by
\begin{equation}\label{genfunc}
\mathcal{Z}[J]=\!\int\mathcal{D}\phi(x)~e^{-i\!\int\frac{1}{2}\partial_x\phi(x)\partial_x\phi(x)+\frac{\omega^2}{2}\phi(x)^2+J\phi(x)~dx},
\end{equation}
where we have replaced the generic potential $V\{\phi(x)\}$ by the quadratic potential $\frac{\omega^2}{2}\phi(x)^2$.
Upon introducing the propagator $A^{-1}$, where
\begin{equation}
A^{1/2}=-\frac{1}{2}\partial_x^2+\frac{\omega^2}{2},
\end{equation}
and completing the square, we obtain:
\begin{equation}
\mathcal{Z}[J]=\mathscr{N}\!\int\mathcal{D}\phi~ e^{-\!\int\!\int\frac{1}{4}J(x)A(x-x')^{-1}J(x')~dx~dx'}.
\end{equation}
The two-point correlation function is then
\begin{equation}
\left<\phi(\xi)\phi(\eta)\right>=\frac{\delta}{\delta J(\xi)}\frac{\delta}{\delta J(\eta)}\mathcal{Z}[J]=-\frac{1}{4}A^{-1}(\xi-\eta).
\end{equation}
The four-point correlator is given by
\begin{align}
\left<\phi(\xi)\phi(\eta)\phi(\zeta)\phi(\theta)\right>&=(-\frac{1}{4})^2 A^{-1}(\xi-\eta)A^{-1}(\zeta-\theta)\nonumber\\
& \quad{} +-\frac{1}{4})^2 A^{-1}(\xi-\zeta)A^{-1}(\eta-\theta)+(-\frac{1}{4})^2 A^{-1}(\xi-\theta)A^{-1}(\eta-\zeta).
\end{align}
If we set $\xi=\eta=\zeta=\theta$, we obtain $\left<\phi(\xi)^4\right>=3\left<\phi(\xi)^2\right>^2$.

\section{Derivation of $\left<\hat{x}^2\right>$ and $\left<\hat{x}^4\right>$}
\subsection{The Transfer Operator}
The discretized path integral we wish to evaluate is
\begin{equation}\label{Z}
\mathcal{Z}=\!\int\limits\prod_{i=1}^N dx_i~\exp\left\{-\sum_{j=1}^N a \left[ \frac{m}{2} \left(\frac{x_{j+1}-x_j}{a}\right)^2+\frac{1}{2}m\omega^2 x_j^2\right]\right\},
\end{equation}
where $a$ is the lattice spacing. We consider $N$ distinct positions $x_1,\ldots,x_N$. The transfer operator $\hat{T}$ is defined by its matrix elements between its position eigenstates,
\begin{equation}\label{opT}
\braket{x'|\hat{T}|x}=\exp\left[-\frac{m}{2a}(x'-x)^2-\frac{m\omega^2 a}{4}(x^2+x'^{2})\right].
\end{equation}
We impose periodic boundary conditions $x_{N+1}=x_1$.  By combining expressions (\ref{Z}) and (\ref{opT}) and making repeated use of the completeness relation
\begin{equation}
1=\ket{x}\bra{x}
\end{equation}
we infer that $\mathcal{Z}=\mathrm{Tr}(\hat{T}^N)$:
\begin{subequations}
\begin{align}
\mathcal{Z}&=\!\int\limits_{x_{N+1}=x_1} dx_N\ldots dx_1~\exp\left[-a\sum_{j=1}^N \frac{m}{2} \left(\frac{x_{j+1}-x_j}{a}\right)^2+\frac{m\omega^2}{2}x_j^2\right] \\
&=\!\int\limits dx_1\ldots dx_N~dx_{N+1}~\delta(x_{N+1}-x_1)~\exp\left[-a\sum_{j=1}^N\frac{m}{2}\left(\frac{x_{j+1}-x_j}{a}\right)^2+\frac{m\omega^2}{4}(x_j^2+x_{j+1}^2)\right]\\
&=\!\int\limits dx_1\ldots dx_N~dx_{N+1}~\delta(x_{N+1}-x_1)\bigg\langle x_{N+1}|\hat{T}|x_N\bigg\rangle \bigg\langle x_N|\hat{T}|x_{N-1}\bigg\rangle \bigg\langle x_{N-1}|\hat{T}|x_{N-2}\bigg\rangle \ldots \bigg\langle x_2|\hat{T}|x_1\bigg\rangle \\
&=\!\int\limits dx_1\ldots dx_N \bigg\langle x_1|\hat{T}|x_N\bigg\rangle \bigg\langle x_N|\hat{T}|x_{N-1}\bigg\rangle \bigg\langle x_{N_1}|\hat{T}|x_{N-2}\bigg\rangle \ldots\bigg\langle x_2|\hat{T}|x_1\bigg\rangle \\
&=\!\int\limits dx_1\bigg\langle x_1|\hat{T}^N|x_1\bigg\rangle 
=\!\int\limits dx \bigg\langle x|\hat{T}^N|x\bigg\rangle
=\mathrm{Tr}(\hat{T}^N). 
\end{align}
\end{subequations}
The commutator of the momentum operator $\hat{p}$ and position operator $\hat{x}$ is defined as:
\begin{equation}\label{commxp}
[\hat{p},\hat{x}]=-i.
\end{equation}
The canonical momentum generates translations:
\begin{equation}
e^{-i\hat{p}\Delta}\ket{x}=\ket{x-\Delta}.
\end{equation}
The use of a (real-valued, $C^{\infty}$) test function makes this apparent. In the position representation the operator $\hat{p}$ takes the form: 
\begin{equation}
\hat{p}=-i\frac{\partial}{\partial x}, 
\end{equation}
and
\begin{subequations}
\begin{align}
e^{-i\hat{p}\Delta}f(x)&=e^{-\Delta d/dx}f(x) \\
&=\left\{1-\Delta \frac{d}{dx}+\frac{1}{2}\Delta^2 \frac{d^2}{dx^2}- \ldots\right\} f(x) x\\
&=f(x)-\Delta \frac{d f(x)}{dx}+\frac{1}{2}\Delta^2 \frac{d^2 f(x)}{dx^2}- \ldots,
\end{align}
\end{subequations}
which is the Taylor series of $f(x-\Delta)$ at the point $x$. Letting $x-x'\equiv \Delta$ be an arbitrary displacement and using
\begin{equation}
e^{-i \hat{p} \Delta}\ket{x}=\ket{x-\Delta}=\ket{x'},
\end{equation}
we can write the operator $\hat{T}$ in terms of $\hat{p}$ and $\hat{x}$.
\begin{subequations}
\begin{align}\label{intT}
\braket{x|\hat{T}|x'}&=\exp\left[-\frac{m}{2a}(x-x')^2-\frac{m\omega^2 a}{4} (x^2+x'^2)\right]  \\
&=\exp\left[-\frac{m\omega^2 a}{4}(x^2+x'^2)\right]\!\int\limits d\Delta~\delta(x-x'-\Delta)\exp\left[-\frac{m \Delta^2}{2a}\right].
\end{align}
\end{subequations}
Using the definition of the inner product,
\begin{equation}
\braket{x'|x-\Delta}\leftrightarrow \delta(x-x'-\Delta),
\end{equation}
the expression (\ref{intT}) is rewritten in bra-ket notation as
\begin{align}
&\exp\left[-\frac{m\omega^2 a}{4}(x^2+x'^2)\right]\!\int\limits d\Delta~\bigg\langle x'\bigg|x-\Delta\bigg\rangle \exp\left[-\frac{m\Delta^2}{2a}\right] \nonumber \\
=\!\int\limits d\Delta \bigg\langle x'\bigg|&\exp\left[-\frac{m\omega^2 a}{4}\hat{x}^2\right]\exp\bigg[-i\hat{p}\Delta\bigg]\exp\left[-\frac{m\Delta^2}{2 a}\right]\exp\left[-\frac{m\omega^2 a}{4}\hat{x}^2\right]\bigg|x\bigg\rangle,
\end{align}
and
\begin{equation}\label{eq0.8}
\hat{T}=\!\int\limits d\Delta~e^{-(m\omega^2 a/4)\hat{x}^2} e^{-i \hat{p} \Delta} e^{-m \Delta^2/2a}e^{-(m\omega^2 a/4)\hat{x}^2}.
\end{equation}
This integral is evaluated by completing the square in the terms containing $\Delta$:
\begin{subequations}
\begin{align}
-{\Delta^2\over2a}-i\hat{p}\Delta&=-{1\over2a}\big(\Delta^2+2ia\hat{p}\big)\\
&=-{1\over2a}\big(\Delta^2+2ia\hat{p}-a^2\hat{p}^2+a^2\hat{p}^2\big)
=-{1\over2a}\big(\Delta+ia\hat{p}\big)^2-{a\over2}\hat{p}^2\, .
\end{align}
\end{subequations}
The integral over $\Delta$ is carried out by changing the integration variable to $s=(\Delta-ia\hat{p})/\sqrt{2a}$, whereupon
\begin{align}
&\! \int_{-\infty}^\infty\exp\bigg(-{\Delta^2\over2a}-i\Delta\hat{p}\bigg)\,d\Delta=e^{-{1\over2}a\hat{p}^2}\!\int_{-\infty}^\infty\exp\bigg[-{1\over2a}\big(\Delta+ia\hat{p}\big)^2\bigg]\,d\Delta\nonumber\\
&=e^{-{1\over2}a\hat{p}^2}\sqrt{2a}\!\int_{-\infty-i a\hat{p}/\sqrt{2a}}^{\infty-i a\hat{p}/\sqrt{2a}} e^{-s^2}\,ds=\sqrt{2\pi a}~e^{-{1\over2}a\hat{p}^2}\, .
\label{eq0.9}
\end{align}
Hence, we obtain
\begin{equation}
\hat{T}=\sqrt{2\pi a/m}~e^{-(m\omega^2 a/4)x^2}e^{-(a/2m)p^2}e^{-(m\omega^2 a/4)x^2}.
\end{equation}

Several comments are in order about the derivation of $\widehat{T}$.  First, the integration over all possible variations $\Delta=x_{i+1}-x_i$ is required, which is a natural result of working on a discretized spatial variable.  Second, in the Gaussian integral in Eq.~(\ref{eq0.9}), we appear to have neglected that the final integral must be evaluated over a contour in the complex plane.   In fact, a careful examination of this issue, which is carried out in Appendix~\ref{secB}, shows that the result in Eq.~(\ref{eq0.9}) is obtained.  Finally, according to the Campbell--Baker--Hausdorff theorem,
\begin{equation}
\widehat{T}=\sqrt{2\pi a}\,e^{-a\widehat{H}+O(a^3)}\, ,
\end{equation}
where $\widehat{H}$ is the Hamiltonian operator for the harmonic oscillator.

%%%%%%%%%%%%%%%%%%%%%%%%%%%%%%%%%%%%%%%%%%%%%%%%%

\section{The Commutator of $\hat{T}$ with $\mathcal{\hat{H}}$}
The commutator of $\hat{x}$ with $\hat{T}$ is calculated by first writing
\begin{subequations}
\begin{align}
[\hat{x},T]&=\sqrt{2\pi a/m}e^{-(m\omega a/4)\hat{x}^2}\left[\hat{x},~e^{-(a/2m)\hat{p}^2}\right]e^{-(m\omega^2 a/4) \hat{x}^2}\\
&=\sqrt{2\pi a/m}e^{-(m\omega a/4)\hat{x}^2}\bigg[\hat{x},~\sum_{n=0}^{\infty}\frac{(-1)^n}{n!}\left(\frac{a}{2m}\right)^n \hat{p}^{2n}\bigg]e^{-(m\omega^2 a/4) \hat{x}^2}.
\end{align}
\end{subequations}
The calculation of $[\hat{x},~\hat{p}^{2n}]$ proceeds as follows:
\begin{subequations}
\begin{align}
n&=0:\quad [\hat{x},~\mathbb{1}]=0 \\
n&=1:\quad [\hat{x},~\hat{p}^2]= \hat{x} \hat{p}^2-\hat{p}^2 \hat{x}=(\hat{p} \hat{x}+i)\hat{p}-\hat{p}\hat{p}\hat{x}=\hat{p}\hat{x}\hat{p}+i\hat{p}-\hat{p}\hat{p}\hat{x}=2i\hat{p} \\
n&=2:\quad [\hat{x},~\hat{p}^4]=\hat{x} \hat{p}^4-\hat{p}^4 \hat{x}=\hat{p} \hat{x} \hat{p}^3-\hat{p} \hat{p}^3 \hat{x}+i \hat{p}=\hat{p}(3 i\hat{p}^2)+i\hat{p}^3=4 i\hat{p}^3.
\end{align}
\end{subequations}

We now use induction to determine the general term in this series:
\begin{equation}\label{commxp1}
[\hat{x},~\hat{p}^k]=i k \hat{p}^{k-1}.
\end{equation}
Observe that the relation (\ref{commxp1}) is true for $k=1$. Assume that $[\hat{x},~\hat{p}^k]=i k \hat{p}^{k-1}$ for some $k\in\mathbb{N}$; then
\begin{equation}
[\hat{x},~\hat{p}^{k+1}]=\hat{x} \hat{p}^{k+1}-\hat{p}^{k+1}\hat{x}=\hat{p}[\hat{x},~\hat{p}^k]+i \hat{p}^k=i(k+1)\hat{p}^k.
\end{equation}
We shall  use the case $k=2n$. Hence
\begin{subequations}
\begin{align}
\left[\hat{x},~\sum_{n=0}^{\infty} \frac{(-1)^n}{n!}\left(\frac{a}{2m}\right)^n \hat{p}^{2n}\right]&=\sum_{n=0}^{\infty}\frac{(-1)^n}{n!}\left(\frac{a}{2m}\right)^n[\hat{x},~\hat{p}^{2n}] \\
=\sum_{n=1}^{\infty}\frac{(-1)^n}{n!}\left(\frac{a}{2m}\right)^n 2i \hat{p}^{2n-1}
&=-\frac{ia}{m} \hat{p} \sum_{n=0}^{\infty}\frac{(-1)^n}{n!}\left(\frac{a}{2m}\right)^n \hat{p}^{2n} \\
& =-\frac{ia}{m}~\hat{p} e^{-(a/2m)\hat{p}^2}.
\end{align}
\end{subequations}
Thus, we find that
\begin{subequations}
\begin{align}\label{C1}
[\hat{x},~\hat{T}]&=\sqrt{2\pi a/m}~e^{-(m\omega^2 a/4)\hat{x}^2}[\hat{x},~e^{-(a/2m)\hat{p}^2}] e^{-(m\omega^2 a/4)\hat{x}^2} \\
&=-\frac{i a}{m} \sqrt{2\pi a/m}~e^{-(m\omega^2 a/4)\hat{x}^2}e^{-(a/2m)\hat{p}^2} \hat{p}~e^{-(m\omega^2 a/4)\hat{x}^2}\nonumber\\
& \quad \times([\hat{p},~e^{-(m\omega^2 a/4)\hat{x}^2}]+e^{-(m\omega^2 a/4)\hat{x}^2}\hat{p}).
\end{align}
\end{subequations}
Proceeding as before, we arrive at
\begin{equation}\label{commxp2}
[\hat{p},~\hat{x}^l]=-i l \hat{x}^{l-1}.
\end{equation}
Therefore,
\begin{align}\label{C2}
[\hat{p},~e^{-(m\omega^2 a/4) \hat{x}^2}]&=\sum_{n=0}^\infty\frac{(-1)^n}{n!}\left(\frac{m\omega^2 a}{4}\right)^n [\hat{p},~\hat{x}^{2n}] \\
&=-i \sum_{n=1}^\infty\frac{(-1)^n}{n!}\left(\frac{m\omega^2 a}{4}\right)^n 2n~\hat{x}^{2n-1}\\
&=\frac{im\omega^2 a}{2} \hat{x} \sum_{n=1}^\infty\frac{(-1)^n}{n!}\left(\frac{m\omega^2 a}{4}\right)^n \hat{x}^{2n} \\
&=\frac{i m\omega a}{2}\hat{x}e^{-(m\omega^2 a/4) \hat{x}^2}.
\end{align}
By combining Eqs.~(\ref{C1}) and (\ref{C2}), we obtain
\begin{align}
[\hat{x},~\hat{T}]&=-\sqrt{2\pi a/m}\frac{i a}{2m}e^{-(m\omega^2 a/4) \hat{x}^2} e^{-(a/2m)\hat{p}^2} e^{-(m\omega^2 a/4) \hat{x}^2}\frac{i m \omega^2 a}{2}~\hat{x} e^{-(m\omega^2 a/4) \hat{x}^2} \\
&=\hat{T}\left[ \frac{a^2\omega^2}{2}\hat{x}-\frac{i a\hat{p}}{m}\right],
\end{align}
 and conclude that  
\begin{equation}
\hat{x} \hat{T}=\hat{T}\left[ \left(1+\frac{a^2\omega^2}{2}\right)\hat{x}-\frac{i a \hat{p}}{m}\right],
\end{equation}
which is Eq.~(C.10) of Ref.~\onlinecite{Creutz}. 

The calculation of $[\hat{p},~\hat{T}]$ proceeds in an analogous manner. We use the auxiliary results (\ref{commxp1}) and (\ref{commxp2}):
\begin{subequations}
\begin{align}
\hat{p}\hat{T}&=\sqrt{2\pi a/m}~e^{-(m\omega^2 a/4)\hat{x}^2}\left(\frac{i m\omega^2 a}{2} \hat{x}+\hat{p}\right)e^{-a/(2m) \hat{p}^2}e^{-(m\omega^2 a/4)\hat{x}^2} \\
&=\sqrt{2\pi a/m}~e^{-(m\omega^2 a/4)\hat{x}^2}\left[\frac{i m\omega^2 a}{2} \hat{x},~e^{-a/(2m) \hat{p}^2}\right]e^{-(m\omega^2 a/4)\hat{x}^2}\nonumber\\
& \quad +\sqrt{2\pi a/m}~e^{-(m\omega^2 a/4)\hat{x}^2}e^{-a/(2m) \hat{p}^2}\left(\frac{i m\omega^2 a}{2} \hat{x}+\hat{p}\right)e^{-(m\omega^2 a/4)\hat{x}^2} \\
&=\sqrt{2\pi a/m}~e^{-(m\omega^2 a/4)\hat{x}^2}e^{-a/(2m) \hat{p}^2}\left\{\frac{i m\omega^2 a}{2} \hat{x}+\left(\frac{a^2\omega^2}{2}+1\right) \hat{p}\right\}e^{-(m\omega^2 a/4)\hat{x}^2} \\
&=\hat{T}\left\{\left(1+\frac{a^2\omega^2}{2}\right)\hat{p}+i a m\omega^2\left(1+\frac{a^2\omega^2}{4}\right)\hat{x}\right\},
\end{align}
\end{subequations}
which is (C.11) of Creutz and Freedman. \cite{Creutz} These equations can be combined to give
\begin{equation}\label{commfinal}
\left[\frac{\hat{p}^2}{m}+m\omega^2 B\hat{x}^2, \hat{T}\right]=0.
\end{equation}
We verify this result here. Let
\begin{equation}\label{AB}
A^2=1+\frac{a^2\omega^2}{2}, \quad B=\left(1+\frac{a^2\omega^2}{4}\right),
\end{equation}
such that
\begin{subequations}
\begin{align}
\left[\hat{x}^2,\hat{T}\right]&=\hat{T}\left\{ (A^2-1)\hat{x}^2-\frac{i a A}{m}(\hat{x}\hat{p}+\hat{p}\hat{x})-\frac{a^2}{m^2}\hat{p}^2\right\} \\
\left[\hat{p}^2,\hat{T}\right]&=\hat{T}\left\{(A^2-1)\hat{p}^2+i a m \omega^2 A B (\hat{x}\hat{p}+\hat{p}\hat{x})-a^2 m^2 \omega^4 B^2 \hat{x}^2\right\}.
\end{align}
\end{subequations}
We show that $\left[\frac{\hat{p}^2}{m}+m\omega^2 B\hat{x}^2, \hat{T}\right]=0$:
\begin{subequations}
\begin{align}
\left[\frac{\hat{p}^2}{m}+m\omega^2 B\hat{x}^2, \hat{T}\right]&=
(A^2-1)/m\hat{p}^2+ia\omega^2 A B(\hat{x}\hat{p}+\hat{p}\hat{x})-a^2m\omega^4B^2\hat{x}^2\nonumber\\
& \quad +m\omega^2 B (A^2-1)\hat{x}^2-ia\omega^2 A B (\hat{x}\hat{p}+\hat{p}\hat{x})-\frac{a^2\omega^2}B\hat{p}^2/m \\
&=\left\{(A^2-1)-a^2\omega^2 B\right\}\left(\hat{p}^2/m+m\omega^2 B\hat{x}^2\right).  
\end{align}
\end{subequations}
Upon re-inserting the definitions of $A$ and $B$ (see Eq.~(\ref{AB})), we arrive at the desired result:
\begin{equation}
(A^2-1)-a^2\omega^2 B=\left(1+\frac{a^2\omega^2}{2}\right)-1-a^2\omega^2\left(1+\frac{a^2\omega^2}{4}\right).
\end{equation}
and the relation (\ref{commfinal}) is recovered.
Thus the simple harmonic oscillator Hamiltonian 
\begin{equation}
\hat{H}=\frac{\hat{p}^2}{2m}+\frac{1}{2}m\omega^2\left(1+\frac{a^2\omega^2}{4}\right)\hat{x}^2
\end{equation}
and the operator $\hat{T}$ share a basis of eigenstates, where $w$, defined by
\begin{equation}
w^2=\omega^2\left(1+\frac{a^2\omega^2}{4}\right).
\end{equation}
is the effective natural frequency of the oscillator.

%%%%%%%%%%%%%%%%%%%%%%%%%%%%%%%%%%%%%%%%%%%%%%%%%

\section{Ladder Operators, the Transfer Operator, and the Hamiltonian}
We now define the ladder operators:
\begin{subequations}
\begin{align}
\hat{a}&=\frac{1}{\sqrt{2mw}}(\hat{p}-imw \hat{x}) \\
\hat{a}^{\dagger}&=\frac{1}{\sqrt{2mw}}(\hat{p}+imw \hat{x}),
\end{align}
\end{subequations}
 Note that these definitions are non-standard.  The Hamiltonian can be written in terms of these operators by first solving Eq.~(\ref{eq22}) for $\hat{x}$ and $\hat{p}$:
\begin{equation}
\hat{p}=\sqrt{\omega\over2}(\hat{a}+\hat{a}^+)\, ,\qquad \hat{x}=i\sqrt{1\over2\omega}(\hat{a}-\hat{a}^+)\, .
\label{eq36}
\end{equation}
By substituting these expressions and invoking Eq.~(\ref{eq25}), we obtain
\begin{subequations}
\begin{align}
\widehat{H}&={\hat{p}^2\over2}+{\omega^2\hat{x}^2\over2} ={\omega\over4}(\hat{a}^++\hat{a})^2-{\omega\over4}(\hat{a}^+-\hat{a})^2\\
&={\omega\over4}\big\{\big[(\hat{a}^+)^2+\hat{a}\hat{a}^++\hat{a}^+\hat{a}+\hat{a}^2\big]-\big[(\hat{a}^+)^2-\hat{a}\hat{a}^+-\hat{a}^+\hat{a}+\hat{a}^2\big]\big\}\\
&={\omega\over2}(\hat{a}\hat{a}^++\hat{a}^+\hat{a})=\big(\hat{a}^+\hat{a}+\textstyle{1\over2}\big)\omega\, .
\end{align}
\end{subequations}
We also have
\begin{subequations}
\begin{align}
[\widehat{H},\hat{a}^+]&=\big(\hat{a}^+\hat{a}+\textstyle{1\over2}\big)\hat{a}^+\omega -\hat{a}^+\big(\hat{a}^+\hat{a}+\textstyle{1\over2}\big)\omega\\
&=\hat{a}^+(\hat{a}^+\hat{a}+1)\omega-(\hat{a}^+\hat{a}^+\hat{a})\omega=\hat{a}^+\omega\, ,\\
[\widehat{H},\hat{a}]&=\big(\hat{a}^+\hat{a}+\textstyle{1\over2}\big)\hat{a}\omega -\hat{a}\big(\hat{a}^+\hat{a}+\textstyle{1\over2}\big)\omega\\
&=(\hat{a}\hat{a}^+-1)\hat{a}\omega-(\hat{a}^+\hat{a}^+\hat{a})\omega=-\hat{a}\omega\, .
\end{align}
\end{subequations}
From the commutator $[\hat{p},\hat{x}]=-i$ we deduce the commutator $\left[\hat{a},\hat{a}^{\dagger}\right]=1$.
\begin{subequations}
\begin{align}
\left[\hat{a},\hat{a}^{\dagger}\right]&=\hat{a}\hat{a}^{\dagger}-\hat{a}^{\dagger}\hat{a}\\
&=\frac{1}{2mw}\{(\hat{p}^2+imw[\hat{p}\hat{x}]+w^2\hat{x}^2)-(\hat{p}^2+imw[\hat{x},\hat{p}]+w^2\hat{x})\} \\
&=\frac{1}{2mw}(2imw[\hat{p},\hat{x}])=i[\hat{p},\hat{x}]=1. 
\end{align}
\end{subequations}
Suppose that $\psi$ is an eigenfunction of $\widehat{H}$ with eigenvalue $E$.  Then,
\begin{subequations}
\begin{align}
\widehat{H}(\hat{a}^+\psi)&=(\widehat{H}\hat{a}^+-\hat{a}^+\widehat{H}+\hat{a}^+\widehat{H})\psi\\
&=(\hat{a}^+\omega+\hat{a}^+E)\psi=(E+\omega)(\hat{a}^+\psi)\, ,\\
\widehat{H}(\hat{a}\psi)&=(\widehat{H}\hat{a}-\hat{a}\widehat{H}+\hat{a}\widehat{H})\psi\\
&=([\widehat{H},\hat{a}]+\hat{a}\widehat{H})\psi=(E-\omega)(\hat{a}\psi)\, .
\end{align}
\end{subequations}
These relations motivate the name ``ladder'' operators for $\hat{a}$ and $\hat{a}^+$ and, more specifically, ``raising'' and ``lowering'' operators for $\hat{a}^+$ and $\hat{a}$, respectively.  Thus, $\hat{a}^+$ changes the eigenstate of $\widehat{H}$ to one with an energy increased by $\omega$, while  $\hat{a}$ changes the eigenstate of $\widehat{H}$ to one with an energy decreased by $\omega$. In Appendix~\ref{secC} we show that the algebraic properties of the raising and lowering operators  mandate that the energy eigenvalues $E_n$ of the quantum harmonic oscillators are $E_n=(n+{1\over2})\omega$ for $n=0,1,2,\cdots$. The ladder operators act on a normalized basis of eigenstates $\{\ket{n}\}$
\begin{equation}
\hat{a}\ket{0}=0, \quad \left(\hat{a}^{\dagger}\right)^n\ket{0}=\sqrt{n!}\ket{n},
\end{equation}
such that $\hat{a}^{\dagger}\hat{a}$ becomes the number operator.
\begin{equation}
\hat{a}\ket{n}=\sqrt{n}\ket{n-1}, \quad \hat{a}^{\dagger}\ket{n}=\sqrt{n+1}\ket{n+1}, \quad \hat{a}^{\dagger}\hat{a}\ket{n}=n\ket{n}. 
\end{equation}
We verify Eq.~(C.21) in Ref.~\onlinecite{Creutz}:
\begin{equation}
\hat{a}\hat{T}=\hat{T}\hat{a}\left(1+\frac{a^2\omega^2}{2}-a\omega\left(1+\frac{a^2m\omega^2}{4}\right)^{1/2}\right).
\end{equation}
By using the definition of $\hat{a}$ and the commutators $[\hat{x},\hat{T}]$ and $[\hat{p},\hat{T}]$,  we obtain 
\begin{subequations}
\begin{align}
[\hat{a},\hat{T}]&=\frac{1}{\sqrt{2mw}}\left\{[\hat{p}\hat{T}]-imw[\hat{x},\hat{T}]\right\}\\
&=\frac{\hat{T}}{\sqrt{2mw}}\left\{\frac{a^2\omega^2}{2}\hat{p}+iam\omega^2\left(1+\frac{a^2\omega^2}{4}\right)\hat{x}
-imw\left(\frac{a^2\omega^2}{2}\hat{x}-\frac{ia\hat{p}}{m}\right)\right\}\\
&=\hat{T}\hat{a}\left(\frac{a^2\omega^2}{2}-a w\right)\\
&=\hat{T}\hat{a}\left(\frac{a^2\omega^2}{2}-a\omega\left(1+\frac{a^2\omega^2}{4}\right)^{1/2}\right),
\end{align}
\end{subequations}
where we have used the definition of $w$. Equivalently,
\begin{align}\label{EqR}
\hat{a}\hat{T}&=\hat{T}\hat{a}\left(1+\frac{a^2 \omega^2}{2}-a\omega\left(1+\frac{a^2\omega^2}{4}\right)^{1/2}\right) =\hat{T}\hat{a}R,
\end{align}
where Eq.~(\ref{EqR}) defines the constant $R$. The first equality shows that $0<R<1$, which will be important for the summations carried out  in the following. 

Because $[\mathcal{\hat{H}},~\hat{T}]=0$, the eigenstates $\{\ket{n}\}$ of $\mathcal{\hat{H}}$ diagonalize $\hat{T}$. Let  $\{\lambda_n\}$ be the eigenvalues:
\begin{equation}
\hat{T}\ket{n}=\lambda_n\ket{n}.
\end{equation}
Note that
\begin{align}\label{aT}
\hat{a}\hat{T}\ket{n}&=\lambda_n\hat{a}\ket{n}=\lambda_n\sqrt{n}\ket{n-1}; \\
\hat{T}\hat{a}R\ket{n}&=R\hat{T}\hat{a}\ket{n}=\sqrt{n}R\hat{T}\ket{n-1}=\sqrt{n}R\lambda_{n-1}\ket{n-1}.\label{TaR}
\end{align}
Because $\hat{a}\hat{T}=\hat{T}\hat{a}R$, the  relations (\ref{aT}) and (\ref{TaR}) imply
\begin{align}
(\hat{a}\hat{T}-\hat{T}\hat{a}R)\ket{n}&=\sqrt{n}(\lambda_n-R\lambda_{n-1})\ket{n-1}=0 \\
\lambda_n&=R\lambda_{n-1}\\
R&=\frac{\lambda_n}{\lambda_{n-1}}.
\end{align}
A relation between $R$ and $\mathcal{\hat{H}}/w$ is established via its diagonal elements (using a normalized basis $\{\ket{n}\}$ of eigenstates):
\begin{align}
\braket{n|\hat{T}|n}&=\lambda_n=R^n\lambda_0=R^{n+1/2}R^{-1/2}\lambda_0 \\
\braket{n|\mathcal{\hat{H}}/w|n}&=n+\frac{1}{2}.
\end{align}
Thus the relation between the two operators is
\begin{equation}
\hat{T}=\sqrt{2\pi a}K R^{\mathcal{\hat{H}}/w}, \label{thiseq}
\end{equation}
where $K$ is a normalization constant which will be determined by calculating the trace of each side of Eq.~\eqref{thiseq}.
\begin{equation}\label{TrTR}
\frac{1}{\sqrt{2\pi a/m}}\mathrm{Tr} (\hat{T})=K\sum_{n=0}^{\infty} \braket{n|R^{\hat{a}^{\dagger}\hat{a}+1/2}|n}=K\sum_{n=0}^{\infty} R^{n+1/2}.
\end{equation}
For the right hand side of Eq.~(\ref{TrTR}), we invoke Eq.~(\ref{EqR}). Thus
\begin{equation}
\sum_{n=0}^{\infty}=\dfrac{R^{1/2}}{1-R}
= \dfrac{\dfrac{a\omega}{2}-\left[1-\dfrac{(a\omega)^2}{4}\right]^{1/2}}{-\dfrac{a\omega}{2}+\left[1-\dfrac{(a\omega)^2}{4}\right]^{1/2}}
=-\dfrac{1}{a\omega}.
\end{equation}
For the left hand side of Eq.~(\ref{TrTR}) we find:
\begin{subequations}
\begin{align}
\frac{1}{\sqrt{2\pi a/m}}\mathrm{Tr}~\hat{T}&=\frac{1}{2\pi}\!\int\limits{dp~dx}~ e^{-m\omega^2 a x^2/2}e^{-a p^2/2m}
=\frac{1}{2\pi}\!\int\limits dx~e^{-(m\omega^2 a/2)x^2}\!\int\limits dp~e^{-(a/2)p^2}\\
&=\frac{1}{2\pi}\sqrt{2\pi}\sqrt{2\pi}(m\omega^2 a)^{-1/2} a^{-1/2m}=\frac{1}{a\omega}.
\end{align}
\end{subequations}
We conclude that $K=-1$. The path integral can now be evaluated in terms of $R$, in the diagonal representation of $\hat{T}$ and $\mathcal{\hat{H}}$:
\begin{subequations}
\begin{align}\label{ZinR}
\mathcal{Z}&=\mathrm{Tr}\left(\hat{T}^N\right)=K^N(2\pi a/m)^{N/2}\sum_{n=0}^N\braket{n|R^{\hat{a}^{\dagger}\hat{a}+1/2}|n} \\
&=K^N(2\pi a R/m)^{N/2}\sum_{n=0}^NR^n=\frac{K^N(2\pi a R/m)^{N/2}}{1-R^N}.
\end{align}
\end{subequations}

%%%%%%%%%%%%%%%%%%%%%%%%%%%%%%%%%%%%%%%%%%%%%%%%%

\section{Correlation Functions}
\subsection{Two-point correlation functions $\langle x_i x_j\rangle$}
Correlation functions follow from the representation [see Ref.~\onlinecite{Creutz}, Eq.~(C.28)]:
\begin{equation}\label{Corr}
\braket{x_i x_{i+j}}=\frac{1}{\mathcal{Z}}\mathrm{Tr}(\hat{x}\hat{T}^j\hat{x}\hat{T}^{N-j}).
\end{equation}
 Equation~(\ref{Corr}) can be derived as follows: 
\begin{subequations}
\begin{align}
\mathcal{Z}\braket{x_i x_{i+j}}&=\!\int\limits\limits_{x_{N+1}=x_1} dx_1 \ldots dx_{N+1}\braket{x_{N+1}|\hat{T}|x_N}\nonumber\\
& \quad \ldots  \times \braket{x_{i+j+1}|\hat{T}|x_{i+j}}\hat{x}\braket{x_{i+j}|\hat{T}|x_{i+j-1}}
\ldots\braket{x_{i+1}|\hat{T}|x_j}\hat{x}\braket{x_i|\hat{T}|x_{i-1}}\ldots\braket{x_2|\hat{T}|x_1}\\
&=\!\int\limits\limits_{x_{N}=x_0} dx_0\ldots dx_{N}\braket{x_{N}|\hat{T}|x_N-1} \ldots \braket{x_{i+j+1}|\hat{T}|x_{i+j}}\hat{x}\braket{x_{i+j}|\hat{T}|x_{i+j-1}}\nonumber\\
& \quad \ldots \times\braket{x_{i+1}|\hat{T}|x_j}\hat{x}\braket{x_i|\hat{T}|x_{i-1}} \ldots\braket{x_1|\hat{T}|x_0}\\
&=\mathrm{Tr}\left\{\hat{T}^{N-(i+j)}\hat{x}\hat{T}^j\hat{x}\hat{T}^i\right\}=\mathrm{Tr}\left\{\hat{x}\hat{T}^j\hat{x}\hat{T}^i\hat{T}^{N-(i+j)}\right\}=\mathrm{Tr}\left\{\hat{x}\hat{T}^j\hat{x}\hat{T}^{N-j}\right\},
\end{align}
\end{subequations}
where we have used the cyclic property of the trace. Next, we show that
\begin{equation}\label{Corrij}
\frac{1}{\mathcal{Z}}\mathrm{Tr}(\hat{x}\hat{T}^j\hat{x}\hat{T}^{N-j})=\frac{R^j+R^{N-j}}{2mw}.
\end{equation}
Armed with Eq.~(\ref{eq36}), which expresses the operator $\hat{x}$ in terms of $\hat{a}$ and $\hat{a}^{\dagger}$, we work out the right hand side of Eq.~(\ref{Corrij}):
\begin{subequations}
\begin{align}\label{summsCorrij}
&\left(\frac{1}{\sqrt{2\pi a/m}K}\right)^N\mathrm{Tr}(\hat{x}\hat{T}^j\hat{x}\hat{T}^{N-j})=\left(\frac{1}{\sqrt{2\pi a/m}K}\right)^N \sum_{n=0}^{\infty}\braket{n|\hat{x}\hat{T}^j\hat{x}\hat{T}^{N-j}|n} \\
&=\sum_{n=0}^{\infty}\braket{n|\hat{x}R^{(a^{\dagger}a+1/2)j}\hat{x}R^{(a^{\dagger}a+1/2)(N-j)}|n}
=R^{N/2}\sum_{n=0}^{\infty}\braket{n|\hat{x}R^{(a^{\dagger}a)j}\hat{x}R^{(a^{\dagger}a)(N-j)}|n} \\
&=R^{N/2}\sum_{n=0}^{\infty}(R^n)^{N-j}
\braket{n|\hat{x}R^{(a^{\dagger}a)j}\hat{x}|n}\\
&=-\frac{R^{N/2}}{2mw}\sum_{n=0}^{\infty}(R^n)^{N-j}
\left(-\sqrt{n}\bra{n-1}+\sqrt{n+1}\bra{n+1}\right)R^{(\hat{a}^{\dagger}\hat{a})j}\nonumber\\
& \quad \times \left(\sqrt{n}\ket{n-1}-\sqrt{n+1}\ket{n+1}\right)\\
&=\frac{R^{N/2}}{2mw}\sum_{n=0}^{\infty}(R^n)^{N-j}\left\{n\braket{n-1|R^{(\hat{a}^{\dagger}\hat{a})j}|n-1}+(n+1)\braket{n+1|R^{(\hat{a}^{(\dagger}a)j}|n+1}\right\}\\
&=\frac{R^{N/2}}{2mw}\sum_{n=0}^{\infty}(R^n)^{N-j}\left\{n R^{(n-1)j}+(n+1)R^{(n+1)j}\right\}\\
&=\frac{R^{N/2}}{2mw}\sum_{n=0}^{\infty}n R^{nN-j}+(n+1) R^{nN+j}.
\end{align}
\end{subequations}
Because $0<R<1$, we can use the identity
\begin{equation}
\sum_{n=0}^{\infty}n\left(R^N\right)^n=\frac{R^N}{(1-R^N)^2}
\end{equation}
to calculate the two sums in Eq.~(\ref{summsCorrij}):
\begin{subequations}
\begin{align}
\sum_{n=0}^{\infty}n R^{nN-j}&=R^{-j}\sum_{n=0}^{\infty}n\left(R^N\right)^n=\frac{R^{N-j}}{(1-R^N)^2}\\
\sum_{n=0}^{\infty}(n+1) R^{nN+j}&=R^j R^{-N}\sum_{n=0}^{\infty}(n+1)R^{(n+1)N}\\
& =R^j R^{-N}\sum_{m=1}^{\infty}m \left(R^N\right)^m =R^j R^{-N}\sum_{m=0}^{\infty}=\frac{R^j}{(1-R^N)^2}.
\end{align}
\end{subequations}
We thus find
\begin{equation}
\left(\frac{1}{\sqrt{2\pi a/m}K}\right)^N\mathrm{Tr}(\hat{x}\hat{T}^j\hat{x}\hat{T}^{N-j})=\frac{R^{N/2}}{2mw}\left\{\frac{R^j+R^{N-j}}{(1-R^N)^2}\right\},
\end{equation}
and finally, making use of   Eq.~(\ref{ZinR}) we recover Eq.~(\ref{Corrij}):
\begin{subequations}
\begin{align}
\braket{x_i x_{i+j}}&=\left(\frac{1}{\sqrt{2\pi a R} K}\right)^N (1-R^N)\frac{R^{N/2}}{2w}\left\{\frac{R^j+R^{N-j}}{(1-R^N)^2}\right\}(\sqrt{2\pi a}K)^N \\
&=\frac{1}{2mw}\left\{\frac{R^j+R^{N-j}}{1-R^N}\right\}.
\end{align}
\end{subequations}
Note that the two-point correlator does not depend on the index $i$.

\subsection{Calculation of the four-point correlation function $\langle\hat{x}^4\rangle$}
Analogously, the four-point correlation function is given by
\begin{equation}
\braket{x_i x_{i+j} x_{i+j+k} x_{i+j+k+l}}=\frac{1}{\mathcal{Z}}\mathrm{Tr}(\hat{T}^{N-(i+j+k+l)}\hat{x}\hat{T}^l\hat{x}\hat{T}^k\hat{x}\hat{T}^j\hat{x}\hat{T}^i),
\end{equation}
with the restriction $i+j+k+l\leq 1$. In particular, if we set the indices $j$, $k$, $l$ to zero ($i$ can be set to zero without loss of generality), we find an expression for the expectation value of the observable $\hat{x}^4$:
\begin{equation}
\braket{\hat{x}^4}=\frac{1}{\mathcal{Z}}\mathrm{Tr}(\hat{x}^4 \hat{T}^N)=\frac{1}{\mathcal{Z}}\mathrm{Tr}(\hat{T}^N\hat{x}^4).
\end{equation} 
As for the derivation Eq.~(\ref{summsCorrij}), we first note that
\begin{equation}\label{Corrx4}
\left(\frac{1}{\sqrt{2\pi a}K}\right)^N\mathrm{Tr}(\hat{T}^N\hat{x}^4)=\sum_{n=0}^{\infty}\braket{n|R^{(\hat{a}^{\dagger}\hat{a}+1/2)N}\hat{x}^4|n}=R^{N/2}\sum_{n=0}^{\infty}\braket{n|R^{(\hat{a}^{\dagger}\hat{a})N}\hat{x}^4|n}.
\end{equation}
To calculate $\hat{x}\ket{n}$, we use that $\hat{x}^4= (\hat{a}-\hat{a}^{\dagger})^4/(2mw)^2$, and
\begin{subequations}
\begin{align}
(\hat{a}-\hat{a}^{\dagger})^4&=\left(\hat{a}^2-\hat{a} \hat{a}^{\dagger}-\hat{a}^{\dagger}\hat{a}+\left(\hat{a}^{\dagger}\right)^2\right)\left(\hat{a}^2-\hat{a} \hat{a}^{\dagger}-\hat{a}^{\dagger}\hat{a}+\left(\hat{a}^{\dagger}\right)^2\right) \\
&=\hat{a}^4-\hat{a}^3 \hat{a}^{\dagger}-\hat{a}^2 \hat{a}^{\dagger}\hat{a}+\hat{a}^2\left(\hat{a}^{\dagger}\right)^2\nonumber\\
& \quad -\hat{a} \hat{a}^{\dagger}\hat{a}^2+\hat{a} \hat{a}^{\dagger} \hat{a} \hat{a}^{\dagger}+\hat{a} \hat{a}^{\dagger}\hat{a}^{\dagger}\hat{a}-\hat{a}\left(\hat{a}^{\dagger}\right)^3\nonumber\\
& \quad -\hat{a}^{\dagger}\hat{a}^3+\hat{a}^{\dagger}\hat{a}^2\hat{a}^{\dagger}+\hat{a}^{\dagger}\hat{a} \hat{a}^{\dagger}\hat{a}-\hat{a}^{\dagger}\hat{a} \left(\hat{a}^{\dagger}\right)^2\nonumber\\
& \quad +\left(\hat{a}^{\dagger}\right)^2 \hat{a}^2-\left(\hat{a}^{\dagger}\right)^2 \hat{a} \hat{a}^{\dagger}-\left(\hat{a}^{\dagger}\right)^3 \hat{a}+\left(\hat{a}^{\dagger}\right)^4.
\end{align}
\end{subequations}
Because the states $\{\ket{n}\}$ are orthogonal, only the six terms in the  following expansion  make a nonzero contribution to the trace: $\hat{a}^2 \left(\hat{a}^{\dagger}\right)^2$, 
$\hat{a} \hat{a}^{\dagger} \hat{a} \hat{a}^{\dagger}$,
 $\hat{a} \hat{a}^{\dagger} \hat{a}^{\dagger} \hat{a}$,
 $\hat{a}^{\dagger} \hat{a} \hat{a} \hat{a}^{\dagger}$,
  $\hat{a}^{\dagger} \hat{a} \hat{a}^{\dagger} \hat{a}$,
and $\left(\hat{a}^{\dagger}\right)^2 \hat{a}^2$. We proceed to normal order these products of operators and act on the ket $\ket{n}$:
\begin{subequations}
\begin{align}
\hat{a} \hat{a} \hat{a}^{\dagger}\hat{a}^{\dagger}&=\hat{a}\{1+\hat{a}^{\dagger}\hat{a}\}=\hat{a} \hat{a}^{\dagger}+\hat{a} \hat{a}^{\dagger} \hat{a} \hat{a}^{\dagger}
=(1+\hat{a}^{\dagger} \hat{a})+\hat{a} \hat{a}^{\dagger}(1+\hat{a}^{\dagger} \hat{a}) \\
&=2(1+\hat{a}^{\dagger}\hat{a})+\hat{a}^{\dagger}\hat{a}+\hat{a}^{\dagger}\hat{a} \hat{a}^{\dagger}\hat{a}
=2+3\hat{a}^{\dagger} \hat{a}+\hat{a}^{\dagger} \hat{a} \hat{a}^{\dagger} \hat{a}\\
(2+ 3 \hat{a}^{\dagger} \hat{a} + \left(\hat{a}^{\dagger} \hat{a}\right)^2)\ket{n}&=(2+3n+n^2)\ket{n}. \\
\hat{a} \hat{a}^{\dagger} \hat{a} \hat{a}^{\dagger}&=\hat{a} \hat{a}^{\dagger}\{1+\hat{a}^{\dagger} \hat{a}\}=\hat{a} \hat{a}^{\dagger}+\hat{a} \hat{a}^{\dagger} \hat{a}^{\dagger} \hat{a} \\
&=(1+\hat{a}^{\dagger} \hat{a})+(1+\hat{a}^{\dagger} \hat{a})\hat{a}^{\dagger} \hat{a}=1+2 \hat{a}^{\dagger} \hat{a}+\hat{a}^{\dagger} \hat{a} \hat{a}^{\dagger} \hat{a} \\
(1+2 \hat{a}^{\dagger} \hat{a}+\hat{a}^{\dagger} \hat{a} \hat{a}^{\dagger} \hat{a})\ket{n}&=(1+2n+n^2)\ket{n}. \\
\hat{a} \hat{a}^{\dagger} \hat{a}^{\dagger} \hat{a}&=(1+\hat{a}^{\dagger} \hat{a})\hat{a}^{\dagger} \hat{a}=\hat{a}^{\dagger} \hat{a}+\hat{a}^{\dagger} \hat{a} \hat{a}^{\dagger} \hat{a} \\
(\hat{a}^{\dagger} \hat{a} + \hat{a}^{\dagger} \hat{a} \hat{a}^{\dagger} \hat{a})\ket{n}&=(n+n^2)\ket{n}. \\
\hat{a}^{\dagger} \hat{a} \hat{a} \hat{a}^{\dagger}&=\hat{a}^{\dagger} \hat{a}(1+\hat{a}^{\dagger} \hat{a})=\hat{a}^{\dagger} \hat{a}+\hat{a}^{\dagger}\hat{a} \hat{a}^{\dagger} \hat{a} \\
(\hat{a}^{\dagger} \hat{a}+\hat{a}^{\dagger} \hat{a} \hat{a}^{\dagger} \hat{a})\ket{n}&=(n+n^2)\ket{n}. \\
\hat{a}^{\dagger} \hat{a} \hat{a}^{\dagger} \hat{a}\ket{n}&=n^2\ket{n}. \\
\hat{a}^{\dagger} \hat{a}^{\dagger} \hat{a} \hat{a} &=\hat{a}^{\dagger}\{\hat{a} \hat{a}^{\dagger}-1\}\hat{a}=\hat{a}^{\dagger} \hat{a} \hat{a}^{\dagger} \hat{a}-\hat{a}^{\dagger} \hat{a}\\
(\hat{a}^{\dagger} \hat{a} \hat{a}^{\dagger} \hat{a}-\hat{a}^{\dagger} \hat{a})\ket{n}&=(n^2-n)\ket{n}.
\end{align}
\end{subequations}
Substituting the normal ordered products into Eq.~(\ref{Corrx4}), the sum becomes
\begin{align}
R^{N/2}\sum_{n=0}^{\infty}\braket{n|R^{(\hat{a}^{\dagger}\hat{a})N}\hat{x}^4|n}
&=\frac{R^{N/2}}{(2 w)^2}\sum_{n=0}^{\infty}\{(2+3n+n^2)+(1+2n+n^2)+2(n+n^2)\nonumber\\
\quad{} + n^2+(n^2-n)\}\braket{n|R^{(\hat{a}^{\dagger}\hat{a})N}|n}
&=\frac{R^{N/2}}{(2 w)^2}\sum_{n=0}^{\infty}\{6n^2+6n+3\}R^{Nn}.
\end{align}
For the three sums we are left with, we shall use the identities ($0<\alpha<1$):
\begin{equation}
\sum_{n=0}^{\infty} \alpha^n=\frac{1}{1-\alpha}, \quad \sum_{n=0}^{\infty}n\alpha^n=\frac{\alpha}{(1-\alpha)^2}, \quad \sum_{n=0}^{\infty} n^2 \alpha^n=\frac{\alpha(1+\alpha)}{(1-\alpha)^3},
\end{equation}
such that
\begin{align}
S_1&\equiv 6 \sum_{n=0}^{\infty} n^2 R^{Nn}=\frac{6 R^N (1+R^N)}{(1-R^N)^3} \\
S_2&\equiv 6\sum_{n=0}^{\infty} n R^{Nn}=\frac{6 R^N}{(1-R^N)^2}\\
S_3&\equiv 3\sum_{n=0}^{\infty} R^{Nn}=\frac{3}{1-R^N},
\end{align}
and
\begin{subequations}
\begin{align}
S_1+S_2+S_3&=\frac{1}{(1-R^N)^3}\left\{6 R^N(1+R^N)+6 R^N(1-R^N)+3(1-R^N)^2\right\} \\
&=\frac{1}{(1-R^N)^3}\left\{3 R^{2N}+6R^N+3\right\}=\frac{3(1+R^N)^2}{(1-R^N)^3},
\end{align}
\end{subequations}
so that finally
\begin{equation}
\frac{1}{\mathcal{Z}}\mathrm{Tr}(\hat{x}^4\hat{T}^N)=\langle\hat{x}^4\rangle=\frac{3}{(2mw)^2}\left(\frac{1+R^N}{1-R^N}\right)^2.
\end{equation}

\section{Pseudocode for the Metropolis update}
\label{Pseudocode}

A sweep produces,  on average,  one attempted update per lattice site  and  requires $3 N_{\tau}$ random numbers. One third is used to specify the ordering in which the sites are visited, one third for the proposed moves, and one third for the Metropolis accept-reject decision. We note that calling random numbers in batches is faster than generating them one by one.

For a given timeslice $\tau$, the proposed value  $x_{\rm new}$  is chosen symmetrically about the present value $x_{\rm old}$. This is the standard recipe to ensure that the algorithm satisfies   detailed balance. 

The meaning of the \emph{if} statement in the  following routine  is summarized as follows. If the action is lowered by the proposed change,  $e^{-s_{\rm new}+s_{\rm old}}>1$, then  the  change is made. 
If  $s_{\rm new}>s_{\rm  old}$,  the use of the random number, uniformly distributed in the interval $[0,1[$, ensures that the proposal is
accepted with the probability  $e^{-s_{\rm  new}+s_{\rm  old}}$.  The random number  $\mbox{randm}[N_{\tau}+i]$  used in the accept/reject step is different from the number  $\mbox{randm}[i]$  used to calculate the proposed new value  $x_{\rm new}$. 

The Mersenne-Twister algorithm \cite{matsumoto98} was used to generate the uniform random numbers. The ideal acceptance rate {\tt idrate} was set to 0.8 at the start of the program.
\begin{center}
\begin{table}[h]
\begin{tabular}{l l}
\emph{Input}: integers $N_{\tau}$, array {\tt path}; real numbers $h$, $m$ and $\omega$. &  \\ 
\emph{Initialize}: real number {\tt accrate=0}. & \\
\emph{Declare}: integers $\tau_{\min}$, $\tau_{\rm plu}$, $i$; &   \\
real numbers $x_{\rm new}$, $s_{\rm old}$, $s_{\rm new}$; &\\
real array $\mbox{randm}(2 N_{\tau})$; integer array {\tt index}$(N_{\tau})$ & \\ 
&  \\
\textbf{for} ($i=0$; $i<N_{\tau}$; $i=i+1$) & specify site visiting order\\
~ ${\tt index[i]}$=floor($N_{\tau}*{\tt getrnd()}$); & \\
\textbf{endfor}& ~ \\
\textbf{for} ($i=0$; $i<2*N_{\tau}$; $i=i+1$) &${\tt getrnd()}$ produces a uniform [0,1[ \\
~${\tt randm[i]} = {\tt getrnd()}$; & random number.   \\
\textbf{endfor}  & \\
\vspace{0.1in}
\textbf{for} ($i=0$; $i<N_{\tau}$; $i=i+1$)  &~\\
~ ~$\tau={\tt index[i]};$ &\\
~ ~$\tau_{\min}=(\tau+N_{\tau}-1)$ modulo $N_{\tau}$; & periodic boundary conditions\\
~ ~$\tau_{\rm plu}=(\tau+1)$ modulo $N_{\tau}$; & \\
~ ~$x_{\rm new}={\tt path}[\tau]+h*({\tt randm}[i]-0.5)$ & proposed new value of ${\tt path}[\tau]$\\
~ ~ $s_{\rm old}~ =\frac{1}{2}m({\tt path}[\tau_{\rm plu}]-{\tt path}[\tau])^2$&\\
~ ~ ~ ~ ~  $~ +\frac{1}{2}m({\tt path}[\tau]-{\tt path}[\tau_{\min}])^2+\frac{1}{2}m\omega^2({\tt path}[\tau])^2;$&current value of the action\\  
~ ~ $s_{\rm new} =\frac{1}{2}m({\tt path}[\tau_{\rm plu}]-x_{\rm new})^2$&\\
~ ~ ~ ~ ~  $~ +\frac{1}{2}m(x_{\rm new}-{\tt path}[\tau_{\min}])^2+\frac{1}{2}m\omega^2(x_{\rm new})^2;$&proposed new value of the action\\  
~ ~ \textbf{if} (${\tt randm}[N_{\tau}+i]<\exp(-s_{\rm new}+s_{\rm old})$)& \\ 
~ ~ ~ ~ ${\tt path}[\tau]=x_{\rm new}$;& build in accepted $x_{\rm new}$\\
~ ~ ~ ~ ${\tt accrate=accrate}+1/N_{\tau}$;&adjustment of acceptance rate\\
~ ~ \textbf{endif}&\\
\textbf{endfor}&\\
$h=h*{\tt accrate}/({\tt idrate})$& adjust target interval for future use\\
\emph{Output}: ${\tt path}$, $h$.
\end{tabular}
\caption{Pseudocode for a Metropolis sweep.}
\end{table}
\end{center}

Within the first {\tt for~loop} of the  routine  ``specify site visiting order'', a time slice $\tau$ may be visited more than once, while another $\tau$ is not visited at all. On average, however, there is one proposed update per site. After $N_{\mathrm{sep}} \gg 1$ sweeps, the differences in updates between the sites are negligible. Alternatively, this piece of code can be
replaced with a call to the following routine, which fills the array {\tt index} with a random permutation of the indices $0,\ldots,N_{\tau}-1$. 

\begin{center}
\begin{table}[h]
\begin{tabular}{l l}
\emph{Input}: integer $N_{\tau}$. &  \\ 
\emph{Initialize}: integer array $p(N_{\tau})$. &\\
\emph{Declare}: integers $i$, $j$, $k$, {\tt tmp}, real array ${\tt randm}(N_{\tau}-1)$. & ~\\
~\\
\textbf{for} ($i=0$; $i<N_{\tau}$; $i=i+1$) & ~  \\
~ $p[i]$=i; & $p$ contains the indices,   \\
\textbf{endfor}& initially in increasing order.  \\
~\\
\textbf{for} ($i=0$; $i<N_{\tau}-1$; $i=i+1$) &${\tt getrnd()}$ produces a uniform [0,1[ \\
~${\tt randm[i]} ={\tt getrnd()}$; & random number.\\
\textbf{endfor}  &  \\
~\\
\textbf{for} ($j=N_{\tau}-1$; $j\geq 1$; $j=j-1$) &~ \\
~$k={\tt floor(double}(j)*{\tt randm}[j-1])$; & random integer between 0 and $j-1$\\
~${\tt tmp}=p[k]$; $p[k]=p[j]$; $p[j]={\tt tmp}$; & interchange $p[k]$ and $p[j]$\\
\textbf{endfor}  &  \\
\emph{Output}: $p$.
\end{tabular}
\caption{Pseudocode for a permutation of the lattice indices.}
\end{table}
\end{center}

\section{Pseudocode for the jackknife average}
\label{PseudocodeJack}

Let $O[.]$ be a one-dimensional array of length $N$, which contains the measurements $O_i$, $i=1\ldots N$, of the observable $O$. 
An unbiased estimator for the sample mean is $mean=sum/N$, where $sum=\sum_{i=1}^N O[i]$. 
An unbiased estimator is obtained this way,  regardless of a possible autocorrelation within the set of measurements.

In case of uncorrelated data, the statistical uncertainty of the mean is related to the standard deviation of the overall distribution by a factor $1/\sqrt{N}$; \cite{Montvay,Rothe}  that is,  the statistical error of the mean is 
$\sqrt{1/(N(N-1))}\sum_{i=1}^N \left[(O[i]-\mbox{mean})^2 \right]$: see Eq.~(58) in the main text.

The goal of the jackknife procedure is to generalize this  relation  to the case where some autocorrelation is present in the data.  For example,  if the data were only pairwise correlated, we could combine two adjacent measurements, and use the  same relation  with $N\to N/2$.

A first step is to divide the measurements into $N/B$ blocks or  bins,  each holding $B$ adjacent measurements. Naturally, the integer $B$ must divide $N$. The bin size $B$ must be small compared to the total number of
measurements for the jackknife error to be based on a sufficient number of block averages, but larger than the autocorrelation time to ensure that correlation between the blocks is minimal. The user is invited to
monitor the estimate of the statistical uncertainty as a function of $B$; it will reach a plateau once $B$ is in the right ballpark.

The second key idea is to operate on ``inverse blocks;''  that is, on all data but a block of $B$ successive measurements. This is vital if the procedure contains, as an intermediate step,  for example,  an effective mass fit to the data
in $O[.]$. (With $B$ measurements the fit often fails to converge, while with $N-B$ data elements the fit runs smoothly.)

In the following routine  the first {\tt for loop} determines the sample mean. The second calculation consists of an inner and outer loop. The result of the inner loop, {\tt elim}, is $B$ times the sample mean of one block (see Eq.~(59) in the main text). In the outer loop,  the jackknife estimator  is calculated.  The jackknife estimator  is the average over all variables but those in the block under consideration, hence based on $N-B$ measurements.

The final loop determines the variance of the jackknife estimator, from which the jackknife error follows by taking a square root, with an appropriate prefactor.

\begin{center}
\begin{table}[h]
\begin{tabular}{l l}
\emph{Input}: array $O$ of size $N$; &  $O$ contains the data; $B$ is the block size \\ 
integer $B$ which divides $N$. & ~ \\
\emph{Declare}: integers $i$, $j$, $n$, &\\
real numbers {\tt sum}, {\tt mean}, {\tt elim}, & \\ 
${\tt  mean}_{j}$, ${\tt variance}_{j}$, ${\tt error}_{j}$; &$j$ is short for ``jackknife''  \\
array ${\tt estimator}_{j}$ of size $N/B$; & \\ 
~\\
\textbf{if} ($B$ does not divide $N$) &\\
~ Drop the first few elements of $O$ such that &\\
~ $B$ divides the number of remaining elements, &\\
~ which becomes the new value of $N$. &\\
\textbf{endif}&\\
~\\
${\tt sum}=0$; & \\
\textbf{for} ($n=0$; $n<N$; $n=n+1$)  &\\
~${\tt sum}+=O[n]$; &\\
\textbf{endfor}&\\
${\tt mean=sum}/N$ & calculating the sample mean\\
\vspace{0.1in}
\textbf{for} ($i=0$; $i<N/B$; $i=i+1$)  &\\
~~${\tt elim}=0$; & \\
~~\textbf{for}($j=i*B$; $j<i*B+B$; $j=j+1$) &  \\
~~~~${\tt elim}+=O[j]$; & summing over the $i$th block\\
~~\textbf{endfor}&\\
~~${\tt estimator}_{j}[i]=({\tt sum-elim})/(N-B)$ &$i$th estimator is based on all variables\\
~& except the $i$th block\\
\textbf{endfor}&\\
~\\
${\tt variance}_j=0;$\\
\textbf{for}($i=0$; $i<N/B$; $i=i+1$)&\\
~~${\tt variance}_{j}+=(N/B-1)/(N-B)({\tt estimator}_{j}[i]-mean)^2$; & calculating the jackknife variance\\
\textbf{endfor}
~\\
${\tt error}_{j}=\sqrt{{\tt variance}_{j}}$&\\
\emph{Output}: ${\tt mean}_{j}$; ${\tt error}_{j}$.
\end{tabular}
\caption{Pseudocode for the jackknife estimator and its error.}
\end{table}
\end{center}

\section{Fresnel Integrals}

\begin{figure}[b!]
\centering
\includegraphics[width=5cm]{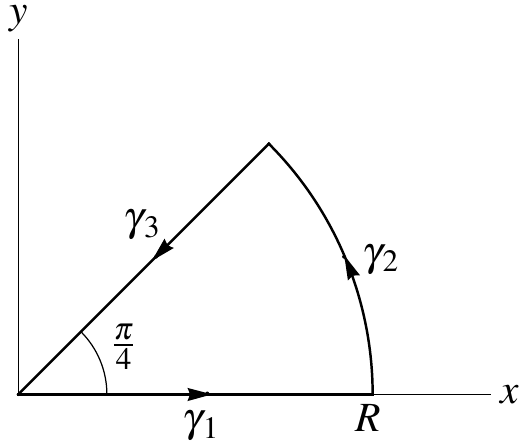}
\caption{The contour in the complex plane used to evaluate the Fresnel integral (\ref{eqA1}).}
\label{figA1}
\end{figure}

The standard complex  Fresnel integral,
\begin{equation}
I=\!\int_0^\infty e^{iz^2}\,dz\, ,
\label{eqA1}
\end{equation}
can be evaluated by contour integration.  The contour, shown in Fig.~\ref{figA1}, consists of three paths: a path $\gamma_1$ along the real line from the origin to $R$, an arc $\gamma_2$ of radius $R$ from the real axis to ${1\over4}\pi$, and a path $\gamma_3$ from the arc back to the origin.  This contour does not enclose any poles for any value of $R$, so the integral over the contour vanishes:
\begin{equation}
\int_{\gamma_1}e^{iz^2}\,dz+\!\int_{\gamma_2}e^{iz^2}\,dz+\!\int_{\gamma_3}e^{iz^2}\,dz=0\, .
\label{eqA2}
\end{equation}
Along $\gamma_1$, $0\le x\le R$, so the integral can be written explicitly as
\begin{equation}
\int_{\gamma_1}e^{iz^2}\,dz=\!\int_0^R e^{ix^2}\,dx\, .
\label{eqA3}
\end{equation}
As $R\to\infty$, this integral becomes the Fresnel integral (\ref{eqA1}).   Along $\gamma_3$, $z=re^{i\pi\over4}$, for $0\le r\le R$, so $dz=e^{i\pi\over4}dr$.  The corresponding integral is
\begin{equation}
\int_{\gamma_3}e^{iz^2}\,dz=e^{i\pi\over4}\!\int_0^R e^{i(re^{i\pi/4})^2}\,dr=e^{i\pi\over4}\!\int_0^Re^{-r^2}\,dr\, ,
\label{eqA4}
\end{equation}
where we have used the fact that $(e^{i\pi\over4})^2=e^{i\pi\over2}=i$.  As $R\to\infty$, this integral becomes a standard Gaussian integral.

Finally, for $\gamma_2$, $z=Re^i\varphi$ for $p\le\varphi\le{1\over4}\pi$, so $dx=iRe^{i\varphi}\,d\varphi$, and the corresponding integral becomes
\begin{equation}
\int_{\gamma_3}e^{iz^2}\,dz=\!\int_0^{\pi\over4}iRe^{i(Re^{i\varphi})^2}\,d\varphi=\!\int_0^{\pi\over4}iRe^{iR^2e^{2i\varphi}}\,d\varphi\, .
\end{equation}
The behavior of this integral as a function of $R$ can be estimated as follows:
\begin{align}
\bigg|\!\int_0^{\pi\over4}iRe^{iR^2e^{2i\varphi}}\,dz\bigg|&\le\!\int_0^{\pi\over4}\big|iRe^{iR^2e^{2i\varphi}}\big|\,dz \\
&=\!\int_0^{\pi\over4}\big|Re^{iR^2(\cos2\varphi+i\sin2\varphi)}\big|\,d\varphi=\!\int_0^{\pi\over4}Re^{-R^2\sin2\varphi}\,d\varphi\, .
\end{align}
Over the interval $0\le\varphi\le{1\over4}\pi$, $\sin2\varphi\ge4\varphi/\pi$, so
\begin{align}
\!\int_0^{\pi\over4}Re^{-R^2\sin2\varphi}\,d\varphi&<\!\int_0^{\pi\over4}Re^{-4R^2\varphi/\pi}\,d\varphi \\
&=-{\pi\over 4R}e^{-4R^2\varphi/\pi}\bigg|_0^{\pi\over4}={\pi\over 4R}\big(1-e^{-R^2}\big)\, ,
\label{eqA7}
\end{align}
which clearly vanishes as $R\to\infty$.

Thus, by combining Eqs.~(\ref{eqA2})--(\ref{eqA4}) and Eq.~(\ref{eqA7}) in the limit $R\to\infty$, we obtain
\begin{equation}
\int_0^\infty e^{ix^2}\,dx=-e^{i\pi\over4}\!\int_0^\infty e^{-r^2}\,dr=-{\sqrt{\pi}e^{i\pi\over4}\over2}\, .
\label{eqA8}
\end{equation}
Thus, the integral in Eq.~(\ref{eq39}) is obtained by taking the complex conjugate of  Eq.~(\ref{eqA8}) and multiplying the result by 2:
\begin{equation}
\int_{-\infty}^\infty e^{-is^2}\,ds=\sqrt{\pi}e^{-i\pi\over4}= {\sqrt{\pi}\over e^{i\pi\over4}}=\sqrt{\pi\over e^{i\pi\over2}}=\sqrt{\pi\over i}\, .
\label{eqA9}
\end{equation}

\section{Gaussian Integrals in the Complex Plane}
\label{secB}

\begin{figure}[b!]
\centering
\includegraphics[width=10cm]{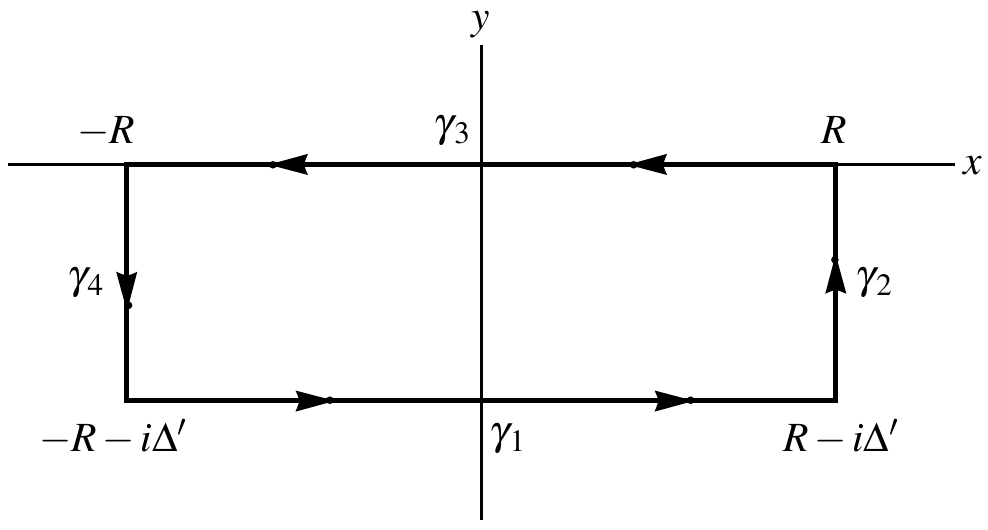}
\vspace{-0.1in}
\caption{The contour in the complex plane used to evaluate the Gaussian integral (\ref{eqB1}).}
\label{figB1}
\end{figure}

The integral in Eq.~(\ref{eq52}) is a Gaussian integral with complex limits:
\begin{equation}
I=\!\int_{-\infty-i\Delta^\prime}^{\infty-i\Delta^\prime}e^{-z^2}\, dz\, ,
\label{eqB1}
\end{equation}
The evaluation of this integral will be carried out over the contour shown in Fig.~\ref{figB1}.  This contour consists of four paths:~a path $\gamma_1$ parallel to the real line from $(-R,-R-i\Delta^\prime)$ to $(R,R-i\Delta^\prime)$, which is continued along path $\gamma_2$ parallel to the imaginary axis $(R,0)$, then along the $x$-axis to $(-R,0)$, and finally along $\gamma_4$ to the original point. This contour does not enclose any poles for any value of $R$, so the integral over the contour vanishes:
\begin{equation}
\int_{\gamma_1}e^{-z^2}\,dz=\!\int_{\gamma_2}e^{-z^2}\,dz+\!\int_{\gamma_3}e^{-z^2}\,dz+\!\int_{\gamma_4}e^{-z^2}\,dz=0\, .
\label{eqB2}
\end{equation}
Along $\gamma_1$, $z=x-i\Delta^\prime$ for $-R\le x\le R$.  Hence, $dz=dx$ and the integral over $\gamma_1$ is
\begin{equation}
\int_{\gamma_1}e^{-z^2}\,dz=\!\int_{-R-i\Delta^\prime}^{R-i\Delta^\prime}e^{-x^2}\,dx\, ,
\end{equation}
which, as $R\to\infty$  becomes the integral in Eq.~(\ref{eqB1}). Along $\gamma_2$, $z=R+iy$, for $-\Delta^\prime\le iy\le 0$.  Thus, $dz=idy$, and the corresponding integral is
\begin{equation}
\int_{\gamma_2}e^{-z^2}\,dz=i\!\int_{-\Delta^\prime}^0e^{-(R+iy)^2}\,dy=i\!\int_0^{\Delta^\prime}e^{-(R-iy)^2}\,dy\, .
\label{eqB4}
\end{equation}
To estimate the magnitude of this integral, we have
\begin{equation}
\bigg|i\!\int_0^{\Delta^\prime}e^{-(R-iy)^2}\,dy\bigg|\le\!\int_0^{\Delta^\prime}\bigg|e^{-(R^2-2iRy-y^2)}\bigg|\,dy=
e^{-R^2}\!\int_0^{\Delta^\prime}e^{y^2}\,dy< \Delta^\prime e^{\Delta^{\prime2}} e^{-R^2}\, ,
\end{equation}
which vanishes as $R\to\infty$.  The integral over $\gamma-3$ is 
\begin{equation}
\int_R^{-R}e^{-x^2}\,dx=-\!\int_{-R}^R e^{-x^2}\,dx\, ,
\end{equation}
which, as $R\to\infty$, becomes a standard Gaussian integral.  Finally,  along $\gamma_4$, $z=R-iy$, for $0\le y\le\Delta^\prime$, so $dz=-idy$ and the integral is
\begin{equation}
\int_{\gamma_2}e^{-z^2}\,dz=-i\!\int_0^{\Delta^\prime}e^{-(R-iy)^2}\,dy\, ,
\end{equation}
which is similar to Eq.~(\ref{eqB4}) and, therefore, also vanishes as $R\to\infty$.  Hence, as $R\to\infty$, Eq.~(\ref{eqB2}) reduces to
\begin{equation}
\int_{-\infty-i\Delta^\prime}^{\infty-i\Delta^\prime}e^{-x^2}\,dx=\!\int_{-\infty}^\infty e^{-x^2}\,dx=\sqrt{\pi}\, .
\label{eqB9}
\end{equation}

\section{Eigenvalues of the Quantum Harmonic Oscillator}
\label{secC}

We have derived three fundamental properties of the raising and lowering operators.  For the purposes of deriving the energy spectrum of the quantum harmonic oscillator, we need only
\begin{equation}
[\hat{a},\hat{a}^+]=1\, ,\qquad \widehat{H}=\big(\hat{a}^+\hat{a}+\textstyle{1\over2}\big)\omega\, .
\label{eqA32}
\end{equation}
Suppose that $\psi$ is an eigenstate of $\widehat{H}$ with eigenvalue $E$:~$\widehat{H}\psi=E\psi$.  Consider the quantity $\hat{a}^+\hat{a}\psi$.  The second of equations (\ref{eqA32}) solved for $\hat{a}^+\hat{a}$ is
\begin{equation}
\hat{a}^+\hat{a}={\widehat{H}\over\omega}-{1\over2}\, ,
\end{equation}
whereupon
\begin{equation}
\hat{a}^+\hat{a}\psi=\bigg({\widehat{H}\over\omega}-{1\over2}\bigg)\psi=\bigg({E\over\omega}-{1\over2}\bigg)\psi\equiv E^\prime\psi\, .
\label{eqA34}
\end{equation}
Operating on both sides of Eq.~\eqref{eqA34}  from the left by $\hat{a}$ yields,
\begin{equation}
\hat{a}(\hat{a}^+\hat{a}\psi)=E^\prime(\hat{a}\psi)\, .
\end{equation}
 By using  the commutation relation in Eq.~(\ref{eqA32}) to write $\hat{a}\hat{a}^+=\hat{a}^+\hat{a}+1$, we find
\begin{equation}
(\hat{a}\hat{a}^+)(\hat{a}\psi)=(\hat{a}^+\hat{a}+1)(\hat{a}\psi)=E^\prime(\hat{a}\psi)\, ,
\end{equation}
or, after a simple rearrangement,
\begin{equation}
\hat{a}^+\hat{a}(\hat{a}\psi)=(E^\prime-1)(\hat{a}\psi)\, .
\end{equation}
Repeating this procedure $k$ times produces
\begin{equation}
\hat{a}^+\hat{a}(\hat{a}^k\psi)=(E^\prime-k)(\hat{a}^k\psi)\, .
\label{eqA37}
\end{equation}
For sufficiently large $k$, we must obtain $\hat{a}^k\psi=0$.  To see this, we multiply Eq.~(\ref{eqA37}) from the left by $(\hat{a}^k\psi)^\dagger$
\begin{align}
(\hat{a}^k\psi)^\dagger \hat{a}^+\hat{a}(\hat{a}^k\psi)&=\big[\hat{a}(\hat{a}^k\psi)\big]^\dagger \hat{a}(\hat{a}^k\psi) \\
&=(\hat{a}^{k+1}\psi)^\dagger (\hat{a}^{k+1}\psi)
=(E^\prime-k)(\hat{a}^k\psi)^\dagger(\hat{a}^k\psi)\, .
\end{align}
 We integrate  over all space (the states considered are bound, so these integrals are finite) and obtain
\begin{equation}
\langle \hat{a}^{k+1}\psi|\hat{a}^{k+1}\psi\rangle=(E^\prime-k)\langle \hat{a}^k\psi|\hat{a}^k\psi\rangle\, .
\end{equation}
Solving for $E^\prime-k$ gives
\begin{equation}
E^\prime-k={|\!|\hat{a}^{k+1}\psi|\!|^2\over|\!|\hat{a}^k\psi|\!|^2}\ge0\, .
\end{equation}
for all $k$.  Thus, if $E^\prime>0$, then $\hat{a}^k\psi$ and $\hat{a}^{k+1}\psi$ are  nonzero.  However, there is a positive integer $n$ such that $a^n\psi\ne0$, but $a^{n+1}\psi=0$; that is, $E^\prime-n=0$.  Therefore, according to Eq.~(\ref{eqA34}), the eigenvalue spectrum of the harmonic oscillator is given by
\begin{equation}
E=\big(n+\textstyle{1\over2}\big)\omega \qquad (n=0,1,2,\cdots)\, .
\end{equation}
 which  has been obtained entirely from the properties of the raising and lowering operators defined in Eq.~(\ref{eq22}); that is, without having to solve the Schr\"odinger equation.


\begin{thebibliography}{47}
\bibitem{f2} Different observables can approach their equilibrium values at different rates.
\bibitem{f3} The local logarithmic slope can also be determined less locally, as long as one stays within the range for which Eq.~(\ref{twoptexp}) holds.
\bibitem{f4} If the error on the error had started to increase at some $\Delta t^{*}<40$, $\Delta t^{*}$ would have defined the cutoff for this parameter set.
\bibitem{UWerr} The Matlab function UWerr.m, Version 6, described in Ref.~\onlinecite{Wolff}, was used to determine the statistical errors of the asymptotic autocorrelation time.
\bibitem{autoremark} If the asymptotic autocorrelation time is fully determined by the term $e^{-t/\tau_0}$, there is no fundamental distinction between the asymptotic and the integrated autocorrelation time, because $\tau_{O,\rm int}\simeq \frac{1}{2}+\int_{1}^{\infty} e^{-\tau/\tau_0}\simeq \int_0^{\infty} e^{-t\tau_0}=\tau_0=\tau_{O,\rm exp}$.
\bibitem{f5} In the continuum limit, the term $1+\tilde{\omega}^2/2$ in Eq.~(\ref{xnewMC}) approaches 1. Thus, in this limit, the trial change $\tilde{x}_i''$ approximates the choice of $\tilde{x}_i'$ by Creutz,\cite{Creutz2} and Brown and Woch.\cite{Brown} We note that the acceptance rate for $\tilde{x}_i''$ appraches  one in the continuum limit, where the kinetic term dominates.
\end{thebibliography}
\end{document}